# Self-Diffusion in a Triple-Defect A-B Binary System:
# Monte Carlo simulation

J. Betlej[1], P. Sowa[1], R. Kozubski[1*], G.E. Murch[2], I.V. Belova[2]

[1]M. Smoluchowski Institute of Physics, Jagiellonian University,
Lojasiewicza 11, 30-348 Krakow, Poland
[2]Centre for Mass and Thermal Transport in Engineering Materials, School of Engineering,
The University of Newcastle, Callaghan 2308, Australia
[*]To whom correspondence should be addressed; E-mail: rafal.kozubski@uj.edu.pl

## Abstract

In this comprehensive and detailed study, vacancy-mediated self-diffusion of A- and B-elements in triple-defect B2-ordered $A_SB_{1-S}$ binaries is simulated by means of a kinetic Monte Carlo (KMC) algorithm involving atomic jumps to nearest-neighbour (nn) and next-nearest-neighbour (nnn) vacancies. The systems are modelled with an Ising-type Hamiltonian with nn and nnn pair interactions completed with migration barriers dependent on local configurations. Self-diffusion is simulated at equilibrium and temperature-dependent vacancy concentrations are generated by means of a Semi Grand Canonical MC (SGCMC) code. The KMC simulations reproduced the phenomena observed experimentally in Ni-Al intermetallics being typical representatives of the 'triple-defect' binaries. In particular, they yielded the characteristic 'V'-shapes of the isothermal concentration dependencies of A- and B-atom diffusivities, as well as the strong enhancement of the B-atom diffusivity in B-rich systems. The atomistic origins of the phenomenon, as well as other features of the simulated self-diffusion such as temperature and composition dependences of tracer correlation factors and activation energies are analyzed in depth in terms of a number of nanoscopic parameters that are able to be tuned and monitored exclusively with atomistic simulations. The roles of equilibrium and kinetic factors in the generation of the observed features are clearly distinguished and elucidated.

## 1. Introduction

The notion of the 'triple defect' was introduced by Wasilewski [1] who originally defined it as a complex of a single A- or B-antisite defect and two nn vacancies in a stoichiometric A-50 at%B system with the B2 superstructure (Fig. 1). Generation of 'triple defects' stems from a substantial difference between the formation energies for A- and B-antisite defects whose consequence is that the system disorders (e.g. due to increasing temperature) by preferentially creating the antisites with lower formation energy. Such a phenomenon is called 'triple-defect disordering' (TDD). It should be noted that in general TDD determines only statistics of the generated defects which may occur without the generation of compact 'triple defects' as defined by Wasilewski [1].

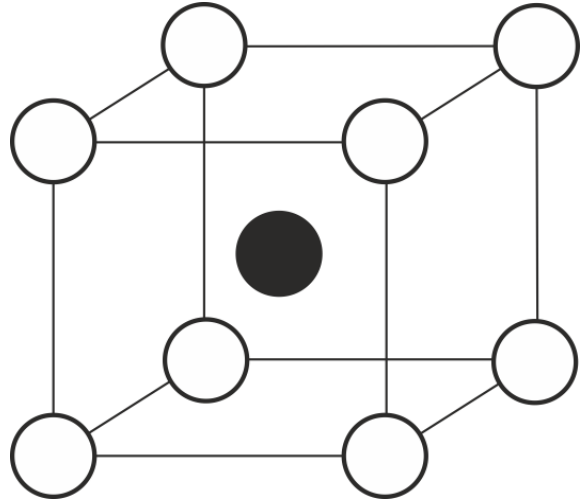

Fig. 1. Scheme of B2-type superstructure: α-sublattice (unfilled circles); β-sublattice (filled circles).

The tendency for TDD implies: (i) a large difference between the A- and B-antisite concentrations; (ii) a large difference between the concentrations of vacancies residing on α- and β-sublattices (the 'home' sublattices of A and B atoms); (iii) an increase of vacancy concentration with decreasing degree of chemical long-range order – i.e. with an increasing concentration of antisite defects. In the extreme case of the exclusive generation of A-antisites, their concentration is equal to one half of the vacancy concentration – i.e. the vacancy concentration strongly increases with decreasing degree of chemical order. In non-stoichiometric binaries the tendency for TDD – i.e. lower formation energy for A-antisites, means that while A-antisites compensate for the deficit of B atoms in A-rich systems, the B-atoms in B-rich systems remain on the β-sublattice and the departure from stoichiometry is compensated by 'structural' α-vacancies. The process of TDD should be contrasted from the so called triple-defect mechanism of diffusion which means atomic migration via specifically correlated atomic jumps mediated by vacancy-pairs [2].

The topic of self or tracer diffusion in stoichiometric and non-stoichiometric A-B B2 intermetallics with the tendency for TDD has been widely investigated. Except for the basic interest in the physical aspects of the phenomenon, the studies were taken up due to technological attraction of the most common TDD alloys (NiAl, FeAl, CoAl, ...). As aluminides, they are successfully applied in industrial manufacturing and in coatings exposed to high temperature in aggressive and corrosive environments.
While the number of theoretical and computational studies is fairly large, experimental works are relatively rare. The main reason for this lies in the major difficulties posed by experimental tracer-diffusion methods that require radioactive isotopes of the constituents. Many TDD alloys have aluminum as the second component which unfortunately lacks suitable radioactive isotopes. The direct tracer-diffusion experiments concerned, therefore the transition-metal components – mostly Ni, whereas the tracer diffusion coefficient of Al has been mainly deduced (approximately) from interdiffusion experiments, the transition-metal self-diffusion coefficient and the thermodynamic factor.

In 1949, Smoluchowski and Burgess [3] measured the tracer diffusion coefficient of Co in NiAl. Co is known for its general tendency to substitute for Ni on the Ni sublattice. Radioactive Co was plated on the NiAl sample and the decrease of activity due to the penetration of cobalt into material was recorded at 1150ºC. The overall shape of the graph showing the self-diffusion coefficient of Co vs. concentration of Ni perfectly resembles one obtained in a more recent experiment by Frank *et al.* [4].

In 1971, Hancock and MacDonnell [5] measured the tracer diffusion coefficient of the radioactive isotope [63]Ni in $Ni_SAl_{1-S}$ polycrystals over the temperature range of 1270 K - 1600 K. The lowest value of diffusion coefficient was observed for the stoichiometric alloy ($S = 0.5$), with a relatively high activation energy of 3.2 eV. It appeared that even a small excess of Al led to a rapid decrease of the activation energy to 1.84 eV for $S = 0.483$. Almost symmetrically, for Ni-rich material, tracer diffusion was also faster than in the stoichiometric system, with a gradual increase of diffusion coefficient and a lowering of the activation energy towards 2.25eV for $S = 0.58$.

These results were critically evaluated by Frank *et al.* [4]. The most significant improvement with respect to the aforementioned study was the almost exclusive use of $Ni_SAl_{1-S}$ monocrystals. Accordingly, the impact of grain boundaries was minimized leading to diffusion coefficients that were smaller by about an order of magnitude. The crucial result of this experiment was that while the Ni-tracer diffusivity systematically rose with $S$ no significant change of Ni tracer diffusivity was observed in the range of $0.468 < S < 0.5$.

Much less information is available concerning Al diffusion. In 1975 Lutze-Birk and Jacobi [6] measured [114]In tracer diffusion in NiAl. Being in the same group as Al in the periodic table, [114]In replaces Al on the Al sublattice. As a function of chemical composition the [114]In-diffusivity showed the characteristic 'V'-shape with the minimum value around the stoichiometric composition.

Indirect estimation of the tracer diffusion coefficients in Ni-Al from interdiffusion experiments has recently been performed by Paul *et al.* [7] and Minamino *et al.* [8]. Fig. 2 shows the results obtained by Paul *et al.* which suggests that when traced in a logarithmic scale versus concentration both Ni and Al tracer diffusivities show again the 'V'-shape with minima around the stoichiometric composition Ni-50at.%Al. Another important feature is the intersection of the $D_{Ni}$ and $D_{Al}$ isotherms at $S < 0.5$.

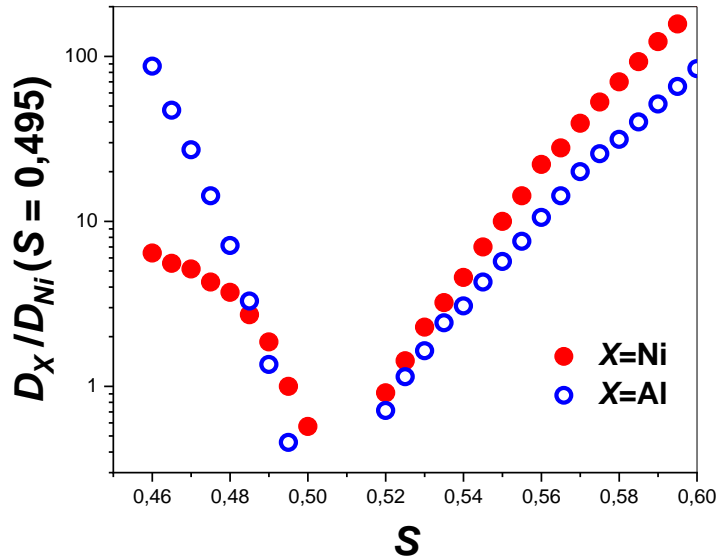

Fig.2. Reduced tracer diffusion coefficients of Ni and Al in Ni$_S$Al$_{1-S}$ intermetallics at 1000$^0$C deduced from interdiffusion experiments. After Paul *et al*. [7]

Predicting rapid growth of the tracer diffusivities of Ni and Al with an increase of Ni and Al content respectively, the results of Paul and Minamino are in good qualitative agreement with those of Frank *et al*. [4] and Hancock and MacDonnell [5]. In addition, the absolute values of the tracer diffusion coefficients calculated for Ni and In [6] are very close to the quantities directly measured in the vicinity of $S = 0.5$. However, the symmetrical growth of the Ni diffusivity with decreasing $S$ (the 'V'-shape) is in clear contrast with the most reliable results of Frank *et al*. [4].

The above experimental results have been widely analyzed in terms of the activation energy of diffusion and possible mechanisms of atomic migration responsible for this energy. Such mechanisms operating in Ni-Al intermetallics at temperatures at which experimental studies are performed are determined by a very high degree of the B2 long-range order (LRO) maintained in these systems up to the melting point of about 1900 K [9]. Krachler *et al*. [10] remarked on the impact of short-range chemical order in Ni-Al resulting in the curved shape of the Arrhenius plots of the measured tracer diffusivities [4]. Mishin *et al*. [11-13] performed extensive studies of the energetics of point defect complexes and migration barriers for various diffusion mechanisms in Ni-Al modelled with interatomic potentials determined within the embedded atom method (EAM). Soule De Bas and Farkas [14] further extended that research by considering complex sequences of 10 and 14 atomic jumps. More recently, Chen *et al*. [15] and Yu *et al*. [16] analyzed the atomic migration barriers in Ni-Al applying either angle-dependent interactions or new EAM potentials fitted to the experimental data. Marino and Carter proposed a more direct computational approach based solely on density functional theory (DFT) and in a series of works [17, 18] evaluated not only the migration barriers and activation energies, but also the diffusion coefficients related to particular mechanisms proposed for Ni tracer diffusion in NiAl.

In 2011 Evteev *et al*. [19] published an interesting paper showing the results on Ni- and Al-self diffusion in NiAl simulated directly by means of Molecular Dynamics. The process was simulated in a layer limited by [110]-oriented free surfaces through which vacancies entered the system from outside and reached an equilibrium concentration. The simulations were performed at a temperature close to the melting point and yielded a Ni-diffusivity ca. 2.5 times higher than the Al diffusivity.

An almost complete computational study of diffusion in NiAl was performed by Xu and Van der Ven [20-22] who combined *ab initio* energy calculations with configurational thermodynamics by means of the Cluster Expansion method. The developed model of Ni-Al covered the equilibrium vacancy thermodynamics with, however, *a priori* assumptions concerning their preferential residence on particular sublattices in the B2 superstructure. Equilibrium thermodynamics of the system including vacancy concentrations was determined by means of the Semi Grand Canonical Monte Carlo (SGCMC) method assuming a zero value of the chemical potential of vacancies. Separately, migration barriers were calculated in the DFT formalism for different jump types and local configurations. Consequently, it was possible to use kinetic Monte Carlo (KMC) to simulate diffusion processes in a system with a well-defined defect concentration and the degree of chemical order. The final results concerning the isothermal concentration dependence of Ni- and Al-diffusivities were in qualitative agreement with the experimental study of Frank *et al.* [4] reproducing the growth of Ni-diffusivity with increasing Ni-concentration in Ni-rich binaries. In agreement with the experimental works of Paul and Minamino [7,8] the same behaviour of Al-diffusivity has been observed. Much less attention has been paid to the Al-rich systems. The results shown in the work concern only one composition ($C_{Ni} \approx 0.47$) and may suggest that the diffusivity growth for both Ni and Al in the Al-rich region is much weaker than that reported experimentally [4-7]. Exploring the tracer-diffusion computations, Xu and Van der Ven evaluated the interdiffusion coefficient for Ni-Al which, however, decreased with growing Al content below $S = 0.5$. Although this clear contradiction with experiment could be attributed to the polycrystalline character of samples analyzed in the works [7, 8], the authors suggested that it rather resulted from incorrect assumptions concerning the equilibrium vacancy concentration.

The present work aims at the determination and detailed analysis of the impact of the tendency for TDD – defined at the beginning of this section, on self-diffusion of the components in B2-ordering A-B binaries. Such studies have been taken up in the past (see e.g. [23]) focusing on the effect of interatomic interactions on the features of diffusion of system component atoms. By adapting in the model relationships between the atomic-jump migration energies yielded by *ab-initio* calculations dedicated to Ni-Al [22], the present study refers specifically to this system. The choice facilitates also the assessment of the simulation findings as the related experimental results with which they might be compared concern almost exclusively Ni-Al. The presented simulations address, therefore, vacancy-mediated atomic migration processes in a B2 superstructure of a TDD system loosely resembling Ni-Al.

By applying a straight forward Ising-type model it is possible to clearly demonstrate the strict correlation between the equilibrium thermodynamics of the system (equilibrium configurations of atoms and vacancies) and the kinetics of self-diffusion. Systems were simulated that represented uniformly a wide range of compositions both in the A-rich and B-rich side of the AB stoichiometry. The approach provides a deep understanding of the diffusion phenomenon which is crucial for any effective development of material technologies.

The paper is organized as follows: The methodology of the study is described in Section 2 clearly pointing at the equilibrium and non-equilibrium (kinetic) aspects of the modelled phenomenon. The model of the simulated A-B binary showing the tendency for TDD and resembling the Ni-Al intermetallic system is presented in Section 3. The results of the study shown in detail in Section 4 are then widely discussed in Section 5. The main conclusions are listed in Section 6.

## 2. Methodology

### 2.1. General remarks

The methodology of the reported study covers two aspects:
- The determination of the temperature and composition dependence of the equilibrium atomic and point-defect configurations in the system.
- The determination of the temperature and composition dependence of self-diffusivities and tracer correlation factors of the system components, as well as their activation energies.

In both cases, Monte Carlo (MC) simulations were performed. Supercells were composed of 25×25×25 unit cells of the B2 superstructure (Fig. 1) – i.e. containing $N = 31250$ lattice sites belonging to equi-numerous α- and β-sublattices and populated with $N_A$ A-atoms, $N_B$ B-atoms and $N_V$ vacancies. 3D periodic boundary condition (PBC) were imposed upon the supercells.

### 2.2. Model for equilibrium configuration of the system:

Of interest are the atomic configurations of a binary A-B system with vacancies. The configurations cover both the distribution of atoms over lattice sites and vacancy concentration and are parameterized by means of the following quantities:
- Concentrations of atoms and vacancies on particular μ-sublattices:

$$C_X^{(\mu)} = \frac{N_X^{(\mu)}}{N} \text{ (X=A,B,V; μ=α,β)} \tag{1}$$

  where $N_X^{(\mu)}$ denotes the number of X species residing on μ-sublattice
- Total concentrations of atoms and vacancies:

$$C_X = C_X^{(\alpha)} + C_X^{(\beta)} \tag{2}$$

- Indicator of the chemical composition:

$$S = \frac{N_A^{(\alpha)} + N_A^{(\beta)}}{N_A^{(\alpha)} + N_A^{(\beta)} + N_B^{(\alpha)} + N_B^{(\beta)}} \tag{3}$$

- Long-range order parameters for atoms and vacancies:

$$\eta_A = \frac{C_A^{(\alpha)} - C_A^{(\beta)}}{C_A^{(\alpha)} + C_A^{(\beta)}}, \ \eta_B = \frac{C_B^{(\beta)} - C_B^{(\alpha)}}{C_B^{(\beta)} + C_B^{(\alpha)}}, \ \eta_V = \frac{C_V^{(\alpha)} - C_V^{(\beta)}}{C_V^{(\alpha)} + C_V^{(\beta)}} \tag{4}$$

- Pair-correlations (short-range order parameters) for atoms and vacancies:

$$C_{XY}^{(\mu\nu)} = \frac{N_{XY}^{(\mu\nu)}}{C_X \cdot N_{tot}^{(\mu\nu)}} \tag{5}$$

  where $N_{XY}^{(\mu\nu)}$ (X,Y =A,B,V; μ,ν = α,β) denotes the number of $X - Y$ nn or nnn pairs with X- and Y-species residing on $\mu -$ and $\nu -$type sublattice sites, respectively; $N_{tot}^{(\mu\nu)} = \sum_{X,Y} N_{XY}^{(\mu\nu)}$

It is important to note that because $N$ is a sum of the numbers of atoms *and vacancies*, a fixed value of $S$ *does not* imply constant values of the component concentrations $C_X$ defined by Eq. (1).

In the equilibrium state corresponding to given external conditions particular configurations appear in the system with a specific probability distribution yielding average values of the above parameters interpreted as observables. By means of the Monte Carlo simulations it is possible to find the equilibrium state of the system by generating a set of configurations showing the equilibrium probability distribution.

The present study was based on the Schapink model for the equilibrium configuration of a multicomponent system with vacancies [24], whose simple version was previously applied by one of the authors [25, 26]. In this approach, a lattice gas A-B-V is treated as a regular *ternary* system – i.e. vacancies are treated strictly as an additional chemical component. The crucial property of the lattice gas (and also the condition for the applicability of the model) is that it shows a miscibility gap with a critical temperature $T_C$ below which it decomposes into two phases: one with $C_V \ll 1$ and another (unrealistic) one with $C_V \approx 1$. Then, the basic assumption of the model is that the lattice-gas phase with $C_V \ll 1$ being in equilibrium with the one with $C_V \approx 1$ is identified with the binary A-B crystal in equilibrium – i.e. the crystal with an equilibrium atomic configuration and equilibrium vacancy concentration.

## 2.3. Search for phase equilibria in the A-B-V lattice gas

Following the idea of Binder *et al.* [27] equilibrium compositions and configurations of the A-B-V lattice gas were determined at fixed temperatures $T$ for arbitrary values of the chemical potentials $\mu_X$ (X=A,B,V). The procedure aimed at finding their values $\mu_X^{(eq)}$ yielding two solutions: one with $C_V \ll 1$ and another with $C_V \approx 1$. Similar to our previous papers (see e.g. [28]) the lattice gas was examined using a standard algorithm of Semi Grand Canonical Monte Carlo (SGCMC) simulations where due to a fixed value of $N$ the system is parameterized by two independent relative chemical potentials defined in the present paper as:

$$\Delta\mu_{AV} = \mu_A - \mu_V$$
$$\Delta\mu_{BV} = \mu_B - \mu_V \tag{6}$$

The SGCMC algorithm works in the following scheme:

(i)     Random choice of a single lattice site occupied by a species 'X' (X = A,B,V);
(ii)    Random choice of a species type 'Y' (Y = A,B,V);
(iii)   Replacement of the species 'X' by the species 'Y' with the Metropolis probability:

$$\Pi_{X\to Y} = \min\left\{1, exp\left[-\frac{\Delta E_{X\to Y} - (\Delta\mu_X - \Delta\mu_Y)}{k_B T}\right]\right\} \tag{7}$$

where $\Delta E_{X\to Y}$ denotes the change of the system configurational energy due to the $X \to Y$ replacement. The quantity $k_B$ denotes the Boltzmann constant. $\Delta E_{X\to Y}$ is evaluated within a particular model of the system implemented with the simulations and depends on the current composition and configuration of the lattice gas.

(iv)  Return to step (i).

Two series of SGCMC runs were performed at each temperature: in series 1 the simulations started with a perfect B2-ordered supercell with $C_A^{(\alpha)} = C_B^{(\beta)} = \frac{1}{2}$; $C_A^{(\beta)} = C_B^{(\alpha)} = C_V = 0$, whereas in the simulations of series 2 the supercell was initially empty ($C_A = C_B = 0$; $C_V = 1$).

The SGCMC simulations run at temperatures below $T_C$ yielded typical $C_V(\Delta\mu_A, \Delta\mu_B)$ isotherms as shown in Fig. 3. The almost cliff-like discontinuity of the $C_V(\Delta\mu_A, \Delta\mu_B)$ surface reflected the coexistence of the vacancy-rich and vacancy-poor phases. The effect showed well-marked hysteresis and thus, exact evaluation of $\Delta\mu_A^{(eq)}$ and $\Delta\mu_B^{(eq)}$ (the white line on the $\Delta\mu_A - \Delta\mu_B$ plane) required an application of some further technique (see e.g. [28,29]). In the present work the technique of thermodynamic integration was chosen (see [28] for detailed description and references).

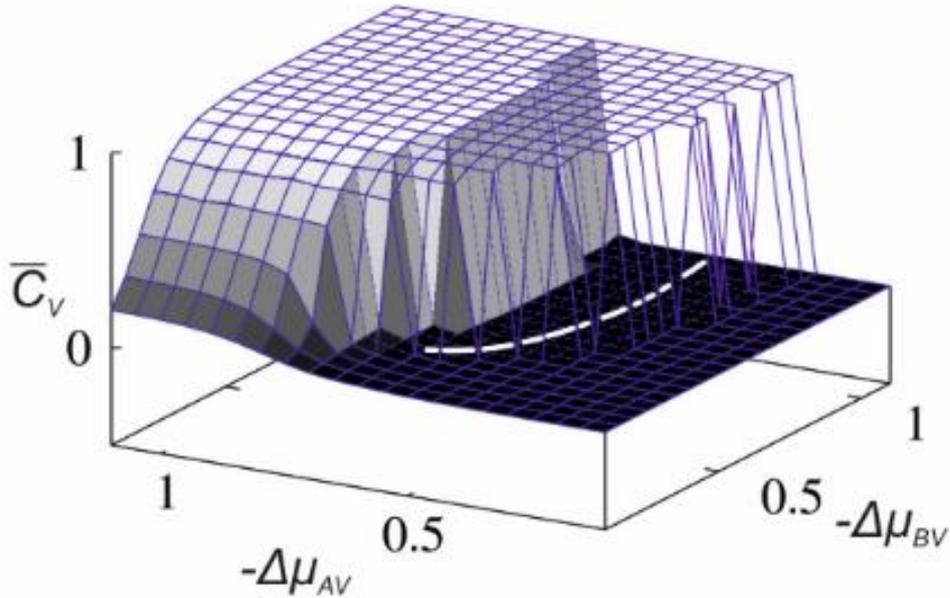

Fig. 3. Typical $C_V(\Delta\mu_A, \Delta\mu_B)$ isotherm with a facet showing the coexistence of the vacancy-rich and vacancy-poor phases. The white line marks the positions of $\Delta\mu_A^{(eq)}$ and $\Delta\mu_B^{(eq)}$.

### 2.4. Model of the vacancy-mediated atomic migration

Vacancy-mediated self-diffusion of A- and B-atoms was simulated at constant temperatures $T$ by means of the standard Residence-Time KMC algorithm [30] in samples with fixed chemical compositions ($S$) and equilibrium vacancy concentrations $C_V$ corresponding to $S$ and $T$ and determined by the SGCMC runs. The initial atomic and vacancy configurations of the samples were generated by former SGCMC runs – i.e. the initial values of $C_X^{(\mu)}$ and $C_{XV}^{(\mu\nu)}$ (X=A,B,V; $\mu,\nu=\alpha,\beta$) were close to the equilibrium (average) ones.

In view of the fact that in most of the previous papers devoted to the modelling of diffusion mechanisms in B2-ordering intermetallics, in particular, in Ni-Al, atomic jumps to both nn and nnn vacancies were considered, the same was implemented in the KMC algorithm applied in the present study.

The probability for an atom X (X =A,B) to jump from the initial $i$ lattice site to a vacancy residing on nn or nnn $j$ lattice site (Fig. 4) is given by:

$$\Pi_{X,i \rightarrow j} = \Pi_0 \times exp\left[-\frac{E_{X,i \rightarrow j}^{(m)}}{k_B T}\right] \tag{8}$$

where: $\Pi_0$ is a pre-exponential factor whose value depends on the jump-attempt frequency of the X-atom and thus is, in general, a function of temperature and the type of jumping atom.

The KMC-time increment of:

$$\Delta t = \tau \times \left\{\sum_{Xij} exp\left[-\frac{E_{X:i \rightarrow j}^{(m)}}{k_B T}\right]\right\}^{-1} \tag{9}$$

is assigned to each executed atomic jump.

$E_{X,i \rightarrow j}^{(m)}$ is the migration barrier for the considered jump and according to Fig. 4:

$$E_{X,i \rightarrow j}^{(m)} = E_{X:i \rightarrow j}^{+} - E_{X:ij} \tag{10}$$

Within the model used in the present paper:

$$E_{X,i \rightarrow j}^{(+)} = \frac{E_{X:i \rightarrow j} + E_{X:j \rightarrow i}}{2} + E_{bar}^{+}(X) \tag{11}$$

where the value of $E_{bar}^{+}(X)$ depends exclusively on the type X of jumping atom.

Consequently,

$$E_{X,i \rightarrow j}^{(m)} = \frac{E_{X:j \rightarrow i} - E_{X:i \rightarrow j}}{2} + E_{bar}^{+}(X) \tag{12}$$

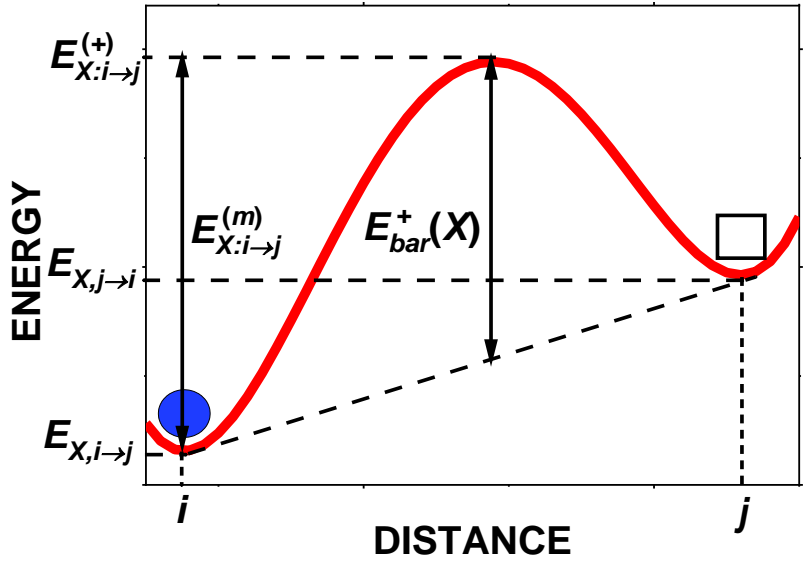

Fig. 4  Scheme of the energy parameterisation of a jump of a X-type atom (blue solid circle) residing on a lattice site i to a vacancy (black open square) residing on a lattice site j.

As was mentioned in our previous works (see e.g. [31,32]) such parameterization of $E_{X,i \to j}^{(m)}$ partially accounts for its dependence on a local configuration around the atom-vacancy pair.

While justification of the negligence of the temperature dependence of $\Pi_0$ (Eq. (1)) was discussed earlier [32], almost equal values of the jump-attempt frequencies reported for Ni and Al-atoms (see e.g. [22,33]) make it reasonable to assume in the present work a constant value of $\Pi_0$ equal to unity both for A- and B-atoms.

Finally, it should be noted that fulfilment of the detailed balance condition [34] by the KMC algorithm guaranteed conservation of the equilibrium configurations of the samples (i.e. maintenance of constant average values of $C_X^{(\mu)}$ and $C_{XV}^{(\mu\nu)}$) all over the KMC simulation runs performed at fixed temperatures.

### 2.5. Evaluation and analysis of diffusivities and correlation factors

The self-diffusion coefficients $D_X$ for X-atoms (X = A, B) were evaluated from the standard Einstein-Smoluchowski relationship (see e.g. [35]):

$$D_X = \lim_{t \to \infty} \left[ \frac{1}{6} \frac{d}{dt} \langle R_X^2(t) \rangle \right] \qquad (13)$$

where $\langle R_X^2(t) \rangle$ denotes the monitored mean-square-distance (MSD) travelled by X-atoms (X = A, B) within the MC-time $t$ – i.e. the value of $R_X^2(t)$ averaged over all X-atoms in the sample. Analysis of the evaluated diffusivities in terms of the dynamics of atomic jumps to vacancies was done within the model of Bakker and colleagues [36] now extended upon atomic jumps to both nn and nnn vacancies. Expression of $\langle R_X^2(t) \rangle$ in Eq. (13) in terms of elementary atomic jumps leads to

$$D_X = \lim_{t \to \infty} \left[ \frac{\{\langle n_X^{(nn)}(t) \rangle \times a_{nn}^2 + \langle n_X^{(nnn)}(t) \rangle \times a_{nnn}^2 \}}{6t} \right] \times f_X^{(corr)} \qquad (14)$$

where $\langle n_X^{(nn)}(t) \rangle$ and $\langle n_X^{(nnn)}(t) \rangle$ denotes the average numbers of nn and nnn jumps performed by an X-atom within the MC-time $t$; $a_{nn}$ and $a_{nnn}$ denote the distances of the nn and nnn jumps, respectively; $f_X^{(corr)}$ denotes the tracer correlation factor given by [37]:

$$f_X^{(corr)} = \lim_{t \to \infty} \left[ \frac{\langle R_X^2(t) \rangle}{\langle n_X^{(nn)}(t) \rangle \times a_{nn}^2 + \langle n_X^{(nnn)}(t) \rangle \times a_{nnn}^2} \right] \qquad (15)$$

The problem is conveniently parameterized with average atomic-jump frequencies $w_{\mu \to \nu}^{(X)}$ defined as average numbers of jumps performed by *one X-atom* from μ-sublattice sites to vacancies residing on ν-sublattice sites (μ,ν = α, β) within a unit KMC-time. Values of $w_{\mu \to \nu}^{(X)}$ are directly determined by counting the particular X-atomic jumps executed within a fixed number of KMC steps and by dividing the number of these jumps by the related KMC time interval and the number $N_X$ of X-atoms present in the supercell. Within the microscopic model [36] they are expressed in terms of the atom-vacancy pair correlations $C_{XV}^{(\mu\nu)}$ (Eq.(5)) and the migration energies associated with the elementary atomic jumps (Eq.(8)):

$$w_{\mu \to \nu}^{(X)}(S,T) = \Pi_0 \times z_{\mu\nu} \times C_{XV}^{(\mu\nu)}(S,T) \times exp \left[ -\frac{\langle E_{X,\mu \to \nu}^{(m)}(S,T) \rangle}{k_B T} \right] \qquad (16)$$

where:

$z_{\mu\nu}$ denotes the number of ν-sublattice sites being nn ($\mu \neq \nu$) or nnn ($\mu = \nu$) of a μ-sublattice site;

$\langle E_{X,\mu \to \nu}^{(m)}(S,T) \rangle$ denotes the average over the migration barriers $E_{X,i \to j}^{(m)}$ associated with the jumps yielding $w_{\mu \to \nu}^{(X)}(S,T)$.

Due to the steady-state character of the simulated self-diffusion – i.e. conservation of the average atomic configuration (i.e. of the values of $\langle C_X^{(\mu)} \rangle$ and $\langle C_{XV}^{(\mu\nu)} \rangle$) guaranteed by the KMC algorithm $w_{\mu \to \nu}^{(X)}$ fulfils the detailed balance condition:

$$w_{\mu \to \nu}^{(X)} = w_{\nu \to \mu}^{(X)} \qquad (17)$$

Eqs. (16) and (17) yield, therefore, a link between the system energetics $\left\{ \langle E_{X,\mu \to \nu}^{(m)} \rangle \right\}$ and the configuration parameters $\left\{ C_{XV}^{(\mu\nu)} \right\}$. Consequently, they also determine the steady-state atomic configuration of the system at a temperature $T$.

Eq. (17) implies that

$$\frac{\langle n_X^{(nn)} \rangle}{t} = 2 w_{\alpha \to \beta}^{(X)}; \quad \frac{\langle n_X^{(nnn)} \rangle}{t} = \left[ w_{\alpha \to \alpha}^{(X)} + w_{\beta \to \beta}^{(X)} \right] \qquad (18)$$

Combination of Eqs. (14) and (18) yields:

$$D_X = \frac{1}{6} \times \left\{ 2 \times w_{\alpha \to \beta}^{(X)} \times a_{nn}^2 + \left[ w_{\alpha \to \alpha}^{(X)} + w_{\beta \to \beta}^{(X)} \right] \times a_{nnn}^2 \right\} \times f_X^{(corr)} \qquad (19)$$

Eq. (19) makes it possible to demonstrate contributions of particular atomic jumps to the observed diffusion coefficients and thus to analyze in such terms the features of the effectively observed self-diffusion.

## 3. Model of the simulated system

### 3.1. Hamiltonian

Applied was an Ising-type model of the B2-ordering binary A-B system with vacancies (and, consequently, of the A-B-V ternary lattice gas) with nearest-neighbour (nn) $\left\{ V_{XY}^{(1)} \right\}$ and next-nearest-neighbour (nnn) $\left\{ V_{XY}^{(2)} \right\}$ pair-interactions between atoms and vacancies: (X,Y = A, B,V). It should be noted that because of the varying composition of the system the SGCMC algorithm involves the total configurational energy (not only the energy of mixing) and therefore, separate evaluation of all the individual pair interaction parameters (not only of the 'ordering energies' $W_{XY}^{(j)} = 2V_{XY}^{(j)} - V_{XX}^{(j)} - V_{YY}^{(j)}$) was required.

Evaluation of the $\left\{ V_{XY}^{(1)} \right\}$ and $\left\{ V_{XY}^{(2)} \right\}$ parameters was based on the following criteria to be fulfilled by the modelled ternary A-B-V lattice gas:

(i)    Ternary miscibility gap with a non-zero critical temperature $T_C$.

(ii)   B2-ordering of the vacancy-poor lattice-gas phase at temperatures below the order-disorder temperature $T_{O-D}$ : $T_{O-D} < T_C$.

(iii)  Tendency for TDD in the vacancy-poor phase – i.e. preferential formation of A-antisite defects. With reference to the earlier remarks (see Section 1) the tendency for TDD was parameterized by the 'triple defect indicator' $TDI$ defined in ref. [36] as $TDI = \frac{N_A^{(\beta)}}{N_V}$. In the stoichiometric AB binary with the tendency for TDD $TDI \approx 1/2$ should hold through a finite temperature range.

As fulfilment of the above criteria determines only the relationships between $\left\{ V_{XY}^{(1)} \right\}$ and $\left\{ V_{XY}^{(2)} \right\}$ assignment of particular values of the pair potentials required an arbitrary evaluation of $T_{O-D}$ in the stoichiometric system with $N_A = N_B$. It should be firmly stressed that by no means did the latter affect meaningful results of the study, which in most cases are presented with relative (reduced) parameters.

The preliminary search for the proper values of $\left\{ V_{XY}^{(1)} \right\}$ and $\left\{ V_{XY}^{(2)} \right\}$ was done by scanning their space and analytically checking the above criteria within the Bragg-Williams approximation (see [38]). As a starting point, the values of $\left\{V_{XY}^{(1)}\right\}$ found in ref. [38] were used. A relationship $V_{XY}^{(2)} = -0.5 \times V_{XY}^{(1)}$ was chosen as an arbitrary assumption. Further adjustment was performed by checking the equilibrium atomic configurations generated by SGCMC simulations for the fulfilment of the criteria listed above. It must be emphasized that no calculations are known by the authors that accounted for interactions with vacancies and no strict reference to literature data was possible.

The final values of $\left\{V_{XY}^{(1)}\right\}$ and $\left\{V_{XY}^{(2)}\right\}$ used in the study are displayed in Table 1.

Table 1. Values of nn ($V_{XY}^{(1)}$) and nnn ($V_{XY}^{(2)}$) pair interaction parameters used in the study.

| X-Y | $V_{XY}^{(1)}$ (eV) | $V_{XY}^{(2)}$ (eV) |
|---|---|---|
| A-A | –0.12 | +0.06 |
| B-B | –0.05 | +0.02 |
| V-V | 0 | 0 |
| A-B | –0.125 | +0.062 |
| A-V | +0.038 | –0.01 |
| B-V | –0.001 | +0.06 |

### 3.2. Migration barriers (saddle-point energies)

For the sake of the studies of atomic migration the extended Ising model was completed with four parameters responsible for atomic migration: $E_{nn,bar}^{+}(A)$, $E_{nn,bar}^{+}(B)$, $E_{nnn,bar}^{+}(A)$ and $E_{nnn,bar}^{+}(B)$ (see Fig. 4 and Eq. (12)). $E_{nn,bar}^{+}(X)$ and $E_{nnn,bar}^{+}(X)$ denote the parameters associated with X-atom jumps to nn and nnn vacancies, respectively.

While the relation $E_{nn,bar}^{+}(A) < E_{nn,bar}^{+}(B)$ followed from the previous work [31], reference to the values of $E_{X,i\rightarrow j}^{(m)}$ determined by *ab initio* calculations of Xu and Van der Ven [22] (Table 2) suggested that $E_{nnn,bar}^{+}(A) > E_{nnn,bar}^{+}(B)$.

Table 2. Values of the migration barriers $E_{X,i\rightarrow i}^{(m)}$ associated with Ni- and Al-atom jumps to nnn vacancies in NiAl – *ab initio* calculations [22].

| $X$ | $E_{X,\alpha\rightarrow\alpha}^{(m)}(eV)$ | $E_{X,\beta\rightarrow\beta}^{(m)}(eV)$ | *Average (eV)* |
|---|---|---|---|
| Ni | 2.76 | 2.05 | 2.41 |
| Al | 2.42 | 1.49 | 1.96 |

Although the evaluation of the $E_{nn,bar}^{+}(X)$ and $E_{nnn,bar}^{+}(X)$ parameters was achieved by respecting the above criteria, the choice of the particular values was arbitrary with a lower limit yielded by the obvious condition of $E_{X,i \rightarrow j}^{(m)} > 0$. Table 3 displays the final values of the $E_{nn,bar}^{+}(X)$ and $E_{nnn,bar}^{+}(X)$ parameters used in the present KMC simulations.

Table 3. Values of the migration-barrier parameters: $E_{nn,bar}^{+}(A)$, $E_{nnn,bar}^{+}(A)$, $E_{nn,bar}^{+}(B)$ and $E_{nnn,bar}^{+}(B)$ used in the study.

| $X$ | $E_{nn,bar}^{+}(X)$ (eV) | $E_{nnn,bar}^{+}(X)$ (eV) |
|---|---|---|
| $A$ | 0.4 | 1.2 |
| $B$ | 0.6 | 0.5 |

## 4. Results

### 4.1. General remarks

Because of the arbitrary evaluation of the energetic parameters of the simulated A-B system, presentation of the simulation results in terms of absolute quantities was generally avoided. The values of particular parameters were, therefore, normalized to selected characteristic values as listed in Table 4.

Table 4. Definitions of normalized (reduced) parameters used in the presentation of the simulation results.

| Reduced Parameter | Symbol | Reference value |
|---|---|---|
| Temperature | $T_{red}^{*}$ | $T_{O-D}(S)$ |
| Temperature | $T_{red}$ | $T_{O-D}^{(max)}$ (Fig.7) |
| Atom-vacancy pair correlation | $\left(C_{XV}^{(\mu\nu)}\right)_{red}$ | $C_{V}(S = 0.5, T_{red}^{*} = 1)$ |
| Tracer diffusivity | $(D_{X})_{red}$ | $D_{A}(S = 0.5, T_{red} = 0.47)$ |
| Activation energy for tracer diffusivity | $\left(E_{A}(D_{X})\right)_{red}$ | $E_{A}(D_{A})(S = 0.5)$ |
| Activation energy for tracer correlation factor | $\left(E_{A}\left(f_{X}^{(corr)}\right)\right)_{red}$ | $E_{A}(D_{A})(S = 0.5)$ |
| Atomic-jump frequency | $\left(w_{\mu\rightarrow\nu}^{(X)}\right)_{red}$ | $w_{\alpha\rightarrow\alpha}^{(A)}(S = 0.5, T_{red} = 0.47)$ |
| Average migration energy | $\langle E_{X,\mu\rightarrow\nu}^{(m)}\rangle_{red}$ | $\langle E_{A,\alpha\rightarrow\alpha}^{(m)}\rangle(S = 0.5, T_{red} = 0.47)$ |

### 4.2. Adequacy and effectiveness of the Schapink model.

Fig. 5 shows an example of the isothermal S-dependence of the relative chemical potentials $\Delta\mu_{AV}^{(eq)}$ and $\Delta\mu_{BV}^{(eq)}$ determined both by directly analyzing the 'map' of Fig. 3 [28] and by calculating their values in equilibrium configurations generated by standard canonical MC simulations [29].

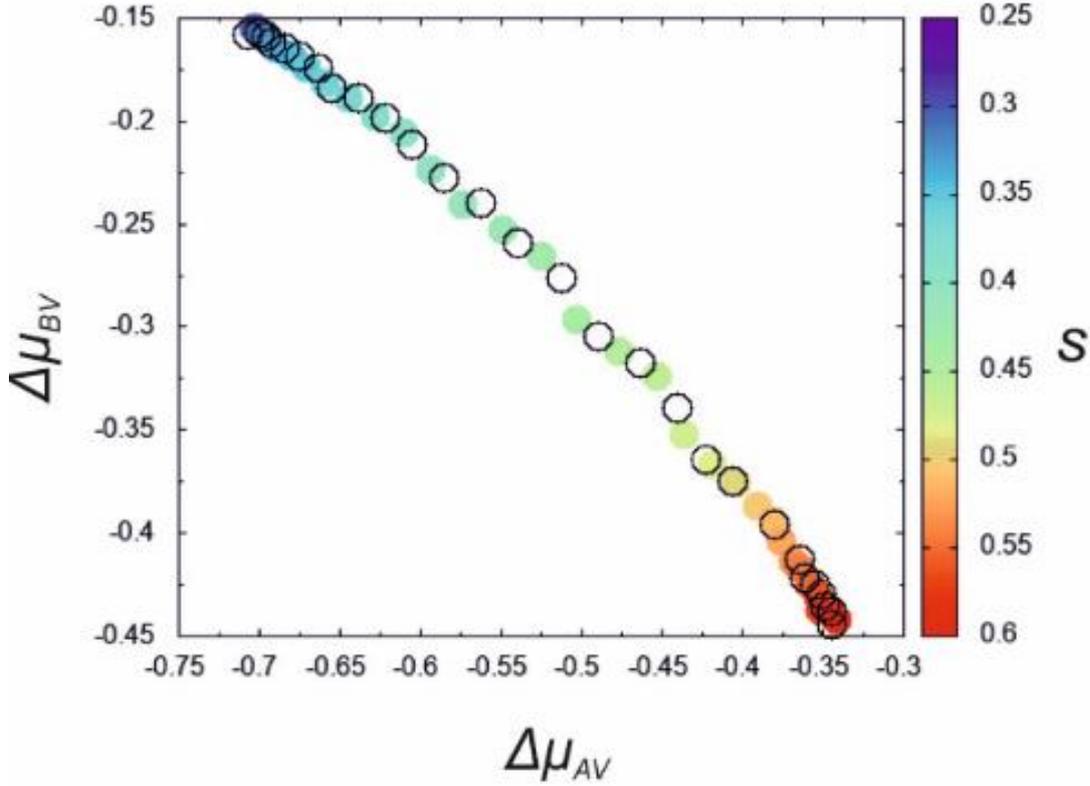

Fig. 5. Relative chemical potentials $\Delta\mu_{X-V}$ corresponding to the phase equilibria in the A-B-V lattice gas at $T_{red} = 0.47$. Coloured solid circles denote the values determined according to ref. [29]; open circles represent the values determined by thermodynamic integration [28].

Almost perfect agreement between the values obtained by both methods indicates the correctness of the present approach.

The SGCMC simulations of the A-B-V lattice gas revealed a miscibility gap, whose several $S = const.$ sections are shown in Fig. 6.

The equilibrium B2-ordered phases with equilibrium vacancy concentrations were analyzed in the range of $0.3 < S < 0.58$ and showed $S$-dependent 'order-disorder' transition temperatures $T_{O-D}$ never exceeding the value of $T_C$ and reaching the maximum value $T_{O-D}^{(max)}$ for $S \approx 0.4$ (Fig.7).

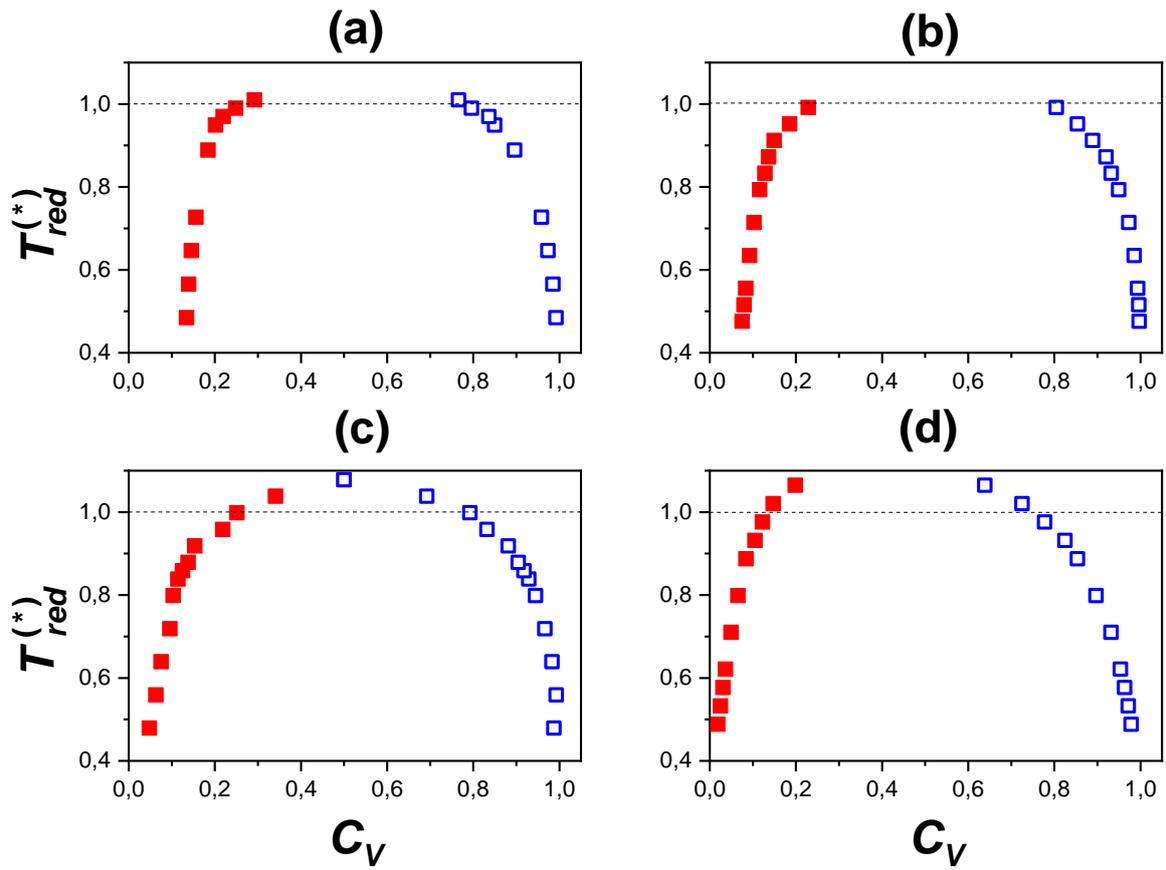

Fig. 6. Sections of the miscibility gap of the A-B-V lattice gas: red solid squares represent the $A_SB_{1-S}$ systems with an equilibrium vacancy concentration: (a) $S = 0.31$; (b) $S = 0.46$; (c) $S = 0.5$; (d) $S = 0.59$.

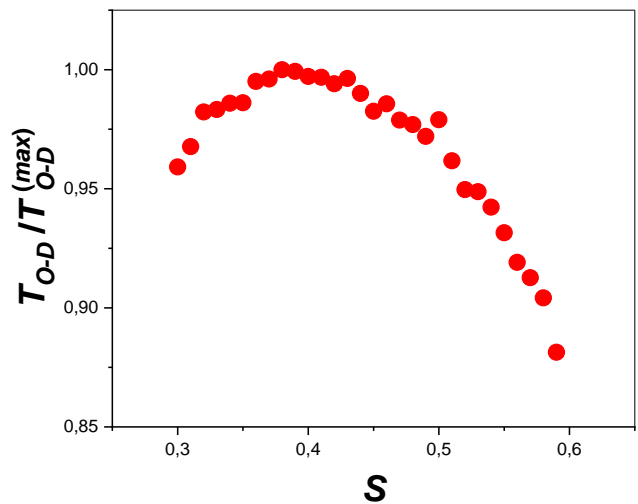

Fig. 7. S-dependence of the reduced 'order-disorder' transition temperature in the simulated A-B binaries with vacancies.

### 4.3. Point defect concentrations and TDD tendency of the system

The SGCMC-generated equilibrium configurations of the system were analyzed in terms of the $T$- and $S$-dependence of point defect concentrations $C_X^{(\mu)}$ (eq.(1)), as well as of atom-vacancy pair correlations $C_{XV}^{(\mu\nu)}$ (Eq.(5)). Arrhenius plots of the parameters were linear in a wide range of temperatures; however, showed well marked curvatures in the vicinity of the 'order-disorder' transition points (see Fig.8 for the plots of $C_V$, $C_V^{(\mu)}$ and $C_{XV}^{(\mu\nu)}$). The effect, obviously following from the increasing temperature dependence of the degree of LRO and SRO, was especially enhanced in the case of $\alpha$-vacancies in B-rich binaries.

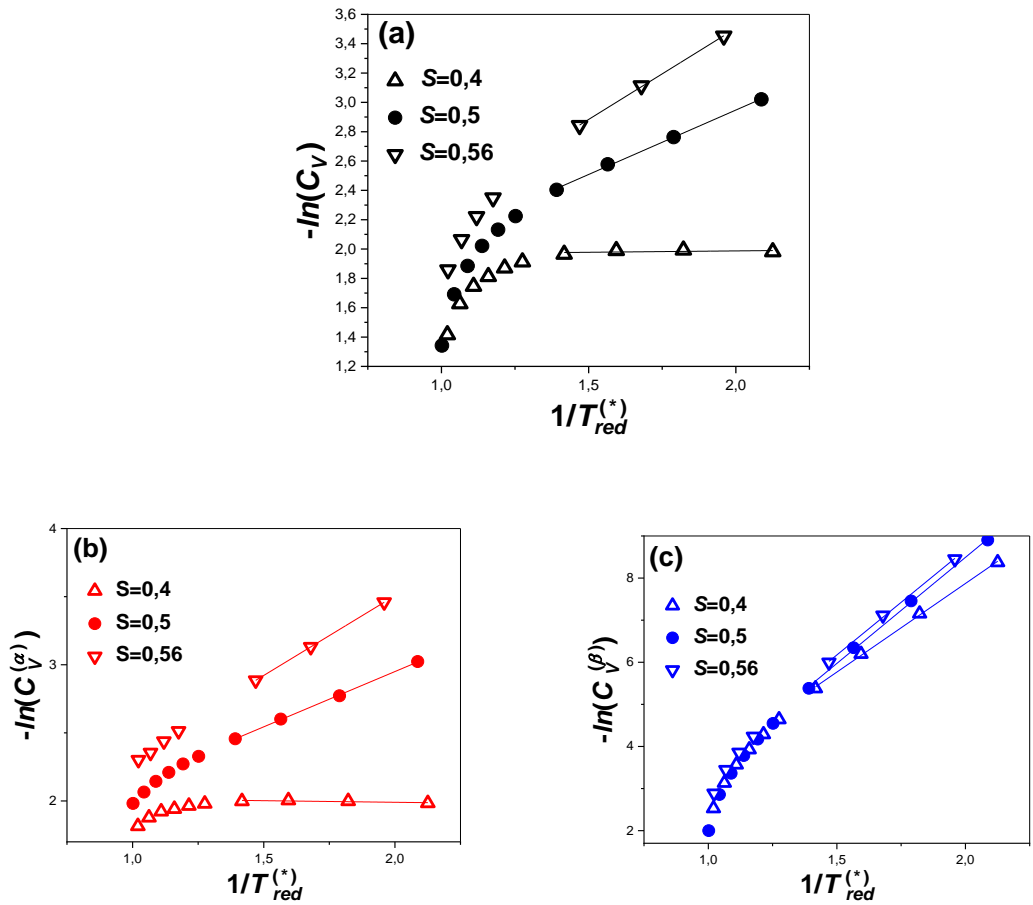

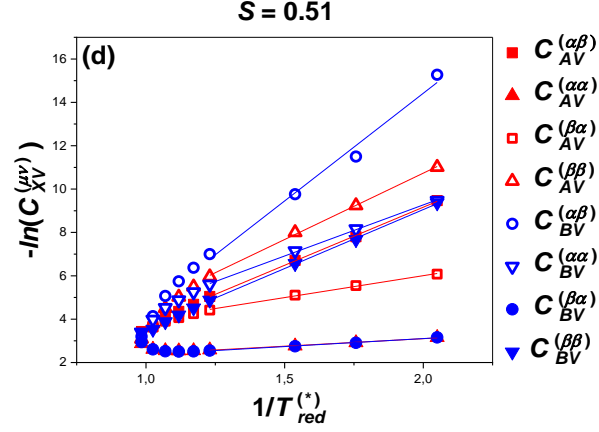

Fig.8. Arrhenius plots of (a) total vacancy concentration $C_V$; (b) $\alpha$-vacancy concentration $C_V^{(\alpha)}$; (c) $\beta$-vacancy concentration $C_V^{(\beta)}$ and (d) atom-vacancy pair correlations $C_{XV}^{(\mu\nu)}$ determined by SGCMC simulations of $A_S B_{1-S}$ binaries.

As briefly described in Section 1, the departure from stoichiometry of a B2 binary A-B system with a tendency for TDD is compensated by structural $\alpha$-vacancies in A-poor binaries ($S < 0.5$) and by structural A-antisites in A-rich ones ($S > 0.5$). The $S$-dependence of the concentrations of the structural point-defects at $T \to 0\,K$ (i.e. at the absence of thermally activated defects) follows from the balance $\sum_{X,\nu} N_X^{(\nu)} = N$ (X=A,B,V; $\nu=\alpha,\beta$) which yields:

(i)    For structural vacancies ($S < 0.5$; $N_A^{(\beta)} = N_B^{(\alpha)} = 0$):

$$N_V^{(\alpha\ struct)} = N_B - N_A \tag{20}$$

$$C_V^{(\alpha\ struct)} = \frac{N_V^{(\alpha\ struct)}}{N} = \frac{N_B - N_A}{N_B + N_A + (N_B - N_A)} = \frac{1 - 2S}{2(1 - S)} \tag{21}$$

(ii)    For structural A-antisites ($S > 0.5$, $C_V = 0$):

$$N_A - N_A^{(\beta\ struct)} = N_B + N_A^{(\beta\ struct)} = \frac{N}{2} \tag{22}$$

$$C_A^{(\beta\ struct)} = \frac{N_A^{(\beta\ struct)}}{N} = \frac{N_A - N_B}{2(N_A + N_B)} = S - \frac{1}{2} \tag{23}$$

Fig. 9 shows the $S$-dependence of the vacancy and antisite concentrations extrapolated to $T \to 0\,K$.

The curves $C_A^{(\beta)}(S, T \to 0K)$ and $C_A^{(\beta\ struct)}(S)$, as well as $C_V^{(\alpha)}(S, T \to 0K)$ and $C_V^{(\alpha\ struct)}(S)$ coincided almost ideally in the range of $S > 0.5$ and $0.45 < S < 0.5$, respectively. This behaviour clearly indicated the triple-defect character of the system. In the range of $S < 0.45$ the curve $C_V^{(\alpha)}(S, T \to 0K)$ deviated, however, from $C_V^{(\alpha\ struct)}(S)$ towards lower values of $C_V^{(\alpha)}$ which was accompanied by an increase of the concentration of B-antisites, appearing already at $S = 0.5$ (see the inset in Fig.9) and contributing to the compensation of the deficit of A-atoms on the α-sublattice. Remarkably, A-antisites definitely absent below $S = 0.5$ re-appeared at $S < 0.4$. The observed effects, especially the decrease of the vacancy concentration below the value of $C_V^{(\alpha\ struct)}$, indicate that the decrease of A-atom concentration caused a decay of the tendency for TDD of the system.

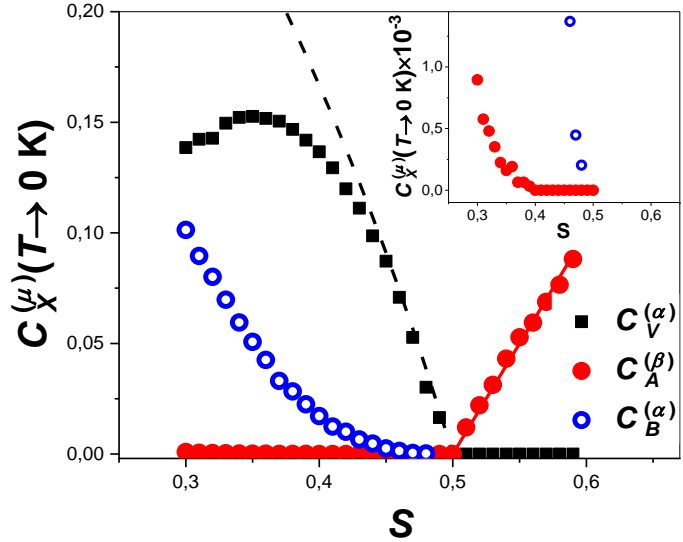

Fig. 9. $S$-dependence of $C_V^{(\alpha)}$, $C_A^{(\beta)}$ and $C_B^{(\alpha)}$ extrapolated to $T \to 0\ K$. The solid red line and the dashed black line denote the $S$-dependences of the concentrations of structural A-antisites and structural A-vacancies, respectively, according to Eqs. (21) and (23).

## 4.4. Temperature and composition dependence of tracer diffusivities of A- and B-atoms

Fig. 10 shows the Arrhenius plots of A- and B-atom diffusivities evaluated by means of Eq. (13) applied to the KMC-time dependences $\langle R_X^2(t) \rangle$ of the MSD yielded by KMC simulations.

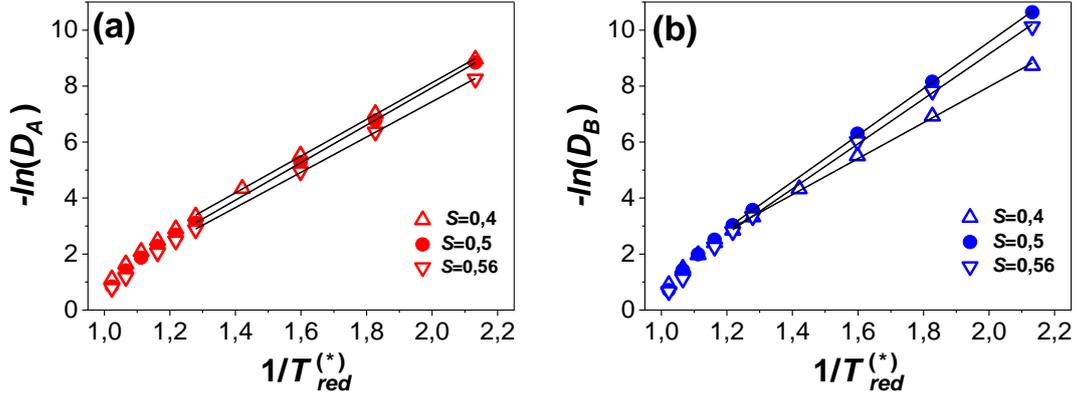

Fig. 10. Examples of the Arrhenius plots of A-atom (a) and B-atom (b) diffusivities determined by KMC simulations of $A_SB_{1-S}$ binaries.

Similarly as in the case of $C_V$, $C_X^{(\mu)}$ and $C_{XV}^{(\mu\nu)}$ the Arrhenius plots of $D_A$ and $D_B$ showed curvatures close to the order-disorder transition point $T_{O-D}$, but the linear segments make it possible to evaluate the activation energies $E_A(D_X)$ for A- and B-atoms self-diffusion (Fig.11).

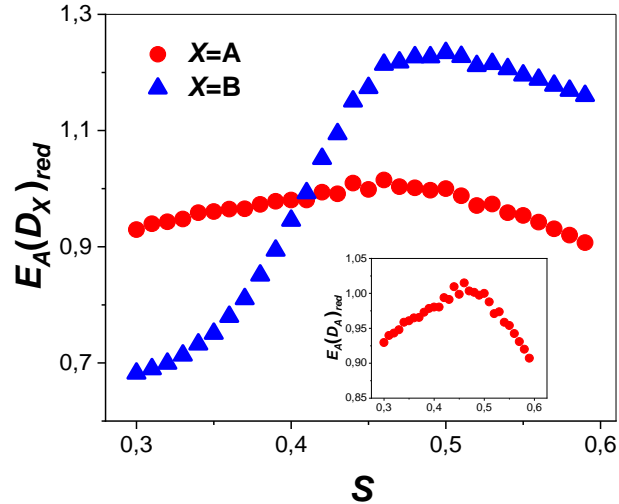

Fig. 11. $S$-dependence of the reduced activation energies $E_A(D_A)_{red}$ and $E_A(D_B)_{red}$.

Both activation energies showed maximum values close to $S = 0.5$. While $E_A(D_A) < E_A(D_B)$ held for $S > 0.4$, a strong decrease of $E_A(D_B)$ with decreasing S caused that the relationship inverted at $S \approx 0.4$. The decrease of $E_A(D_A)$ for $S > 0.5$ means there is qualitative agreement between the reported simulation results and the corresponding experimental data on Ni-tracer diffusion in NiAl [4,5].

Fig. 12 presents the isotherms $(D_A)_{red}(S)$ and $(D_B)_{red}(S)$ corresponding to $T_{red} = 0.47$ and $T_{red} = 0.78$. The curves showed the characteristic asymmetric 'V'-shapes with minima located at $S = 0.43$ and $S = 0.5$, respectively (see the inset in Fig.11a). The 'V'-shape of $(D_B)_{red}(S)$ was definitely more pronounced and clearly visible in a logarithmic scale (Fig. 11c,d). Besides, the curve increased much stronger with *decreasing S* than did $(D_A)_{red}(S)$ with *increasing S*.

As a result, the relationship $D_A > D_B$ observed in the range of A-rich binaries inverted at $S \approx 0.4$ where the $(D_A)_{red}(S)$ and $(D_B)_{red}(S)$ curves intersected. Both features qualitatively corroborate with the experimental results [4,6,7].

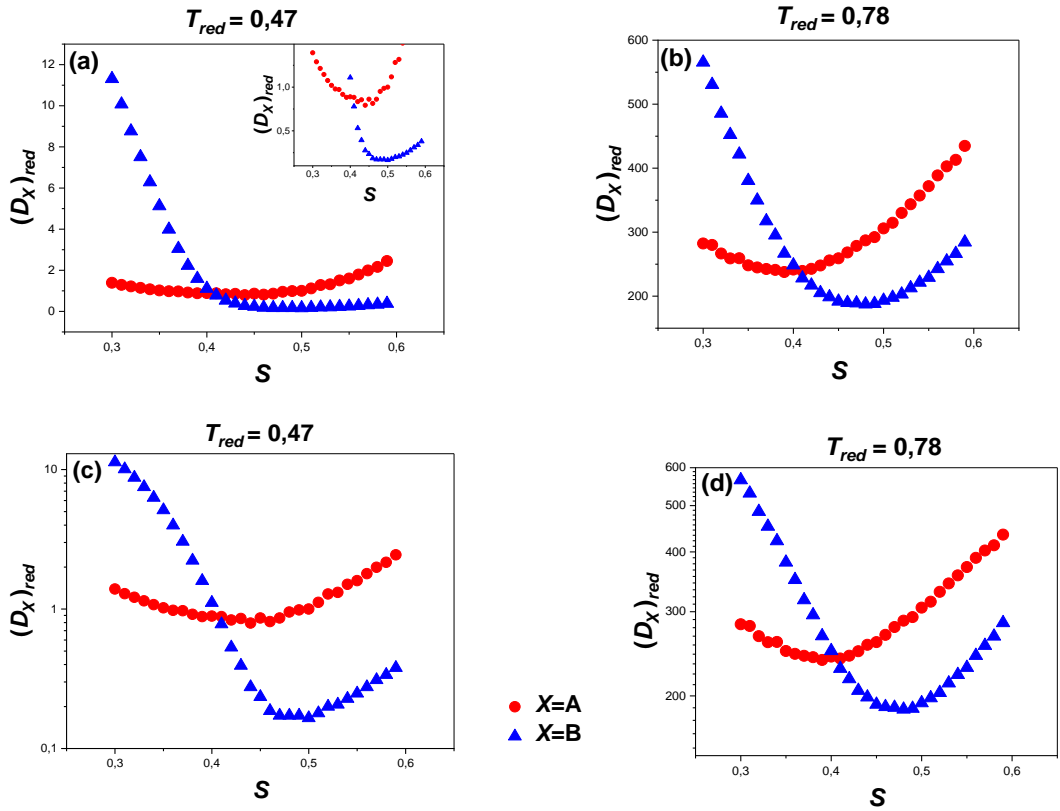

Fig. 12. Examples of the isotherms $(D_A)_{red}$ and $(D_B)_{red}$ traced in linear (a), (b) and in logarithmic (c), (d) scales.

Temperature dependences of the positions of the diffusivity minima and of the intersection of $D_A(S)$ and $D_B(S)$ are displayed in Figs.13a and 13b. While the minima of both $D_A(S)$ and $D_B(S)$ shifted towards lower values of S with increasing temperature (Fig. 13a), the location of $D_A = D_B$ remained at $S \approx 0.4$ in the whole range of $0.47 < T_{red} < 0.8$ (Fig. 13b) – which obviously resulted in $E_A(D_A) = E_A(D_B)$ observed at the same value of $S$ (Fig. 11).

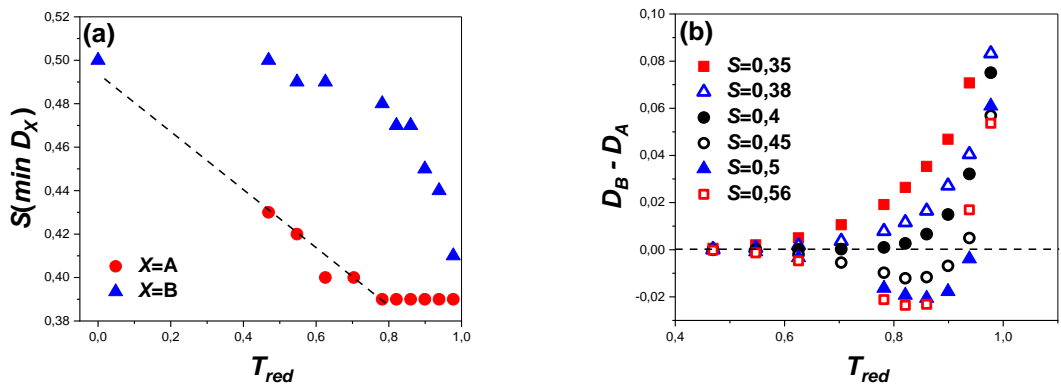

Fig.13. (a) Reduced-temperature dependence of the positions of the minima of $(D_A)_{red}(S)$ and $(D_B)_{red}(S)$; (b) Reduced-temperature dependence of the difference $D_B - D_A$ for selected values od S.

### 4.5. Elucidation of atomistic origins of the features of A- and B-tracer diffusivities

The analysis was based on the atomistic model of self-diffusion and the relationships given by Eqs. (13)-(19) expressing self-diffusion coefficients $D_X(T,S)$ (observables) in terms of tracer correlation factors $f_X^{(corr)}(T,S)$ and atomic-jump frequencies $w_{\mu\to\nu}^{(X)}(T,S)$ which, in turn, depended on atom-vacancy pair correlations $C_{XV}^{(\mu\nu)}(T,S)$ and the average migration barriers $\langle E_{X,\mu\to\nu}^{(m)}\rangle(T,S)$ (Eq. (16)). Each one of the above parameters, as well as its composition- and temperature-dependence was independently evaluable by means of MC simulations.

#### 4.5.1. Tracer correlation factors

Fig. 14 shows the temperature and composition dependences of the tracer correlation factors $f_A^{(corr)}$ and $f_B^{(corr)}$.

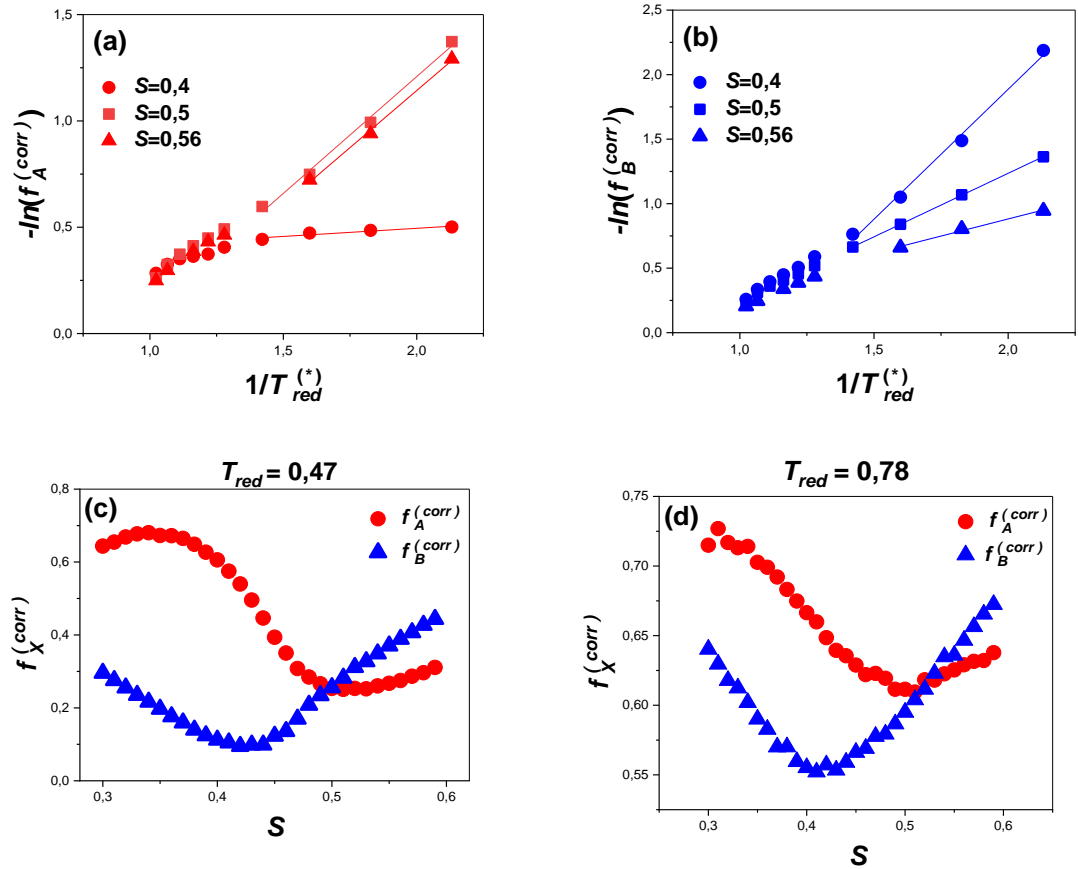

Fig. 14. Tracer correlation factors for A- and B-atom diffusion: Examples of Arrhenius plots of $f_A^{(corr)}$ (a) and $f_B^{(corr)}$ (b); Examples of the isotherms at $T_{red} = 0.47$ (c) and at $T_{red} = 0.78$ (d).

The linear parts of the Arrhenius plots of $f_X^{(corr)}$ yielded effective activation energies $E_A\left(f_X^{(corr)}\right)$ traced in Fig. 15a against $S$. Figs. 15b and 15c show the $S$-dependence of two relationships between $E_A\left(f_X^{(corr)}\right)$ and the total activation energies $E_A(D_X)$ for self-diffusion (Fig. 11): the difference between both activation energies (Fig. 15b) and the contribution of $E_A\left(f_X^{(corr)}\right)$ to $E_A(D_X)$ (Fig.15c).

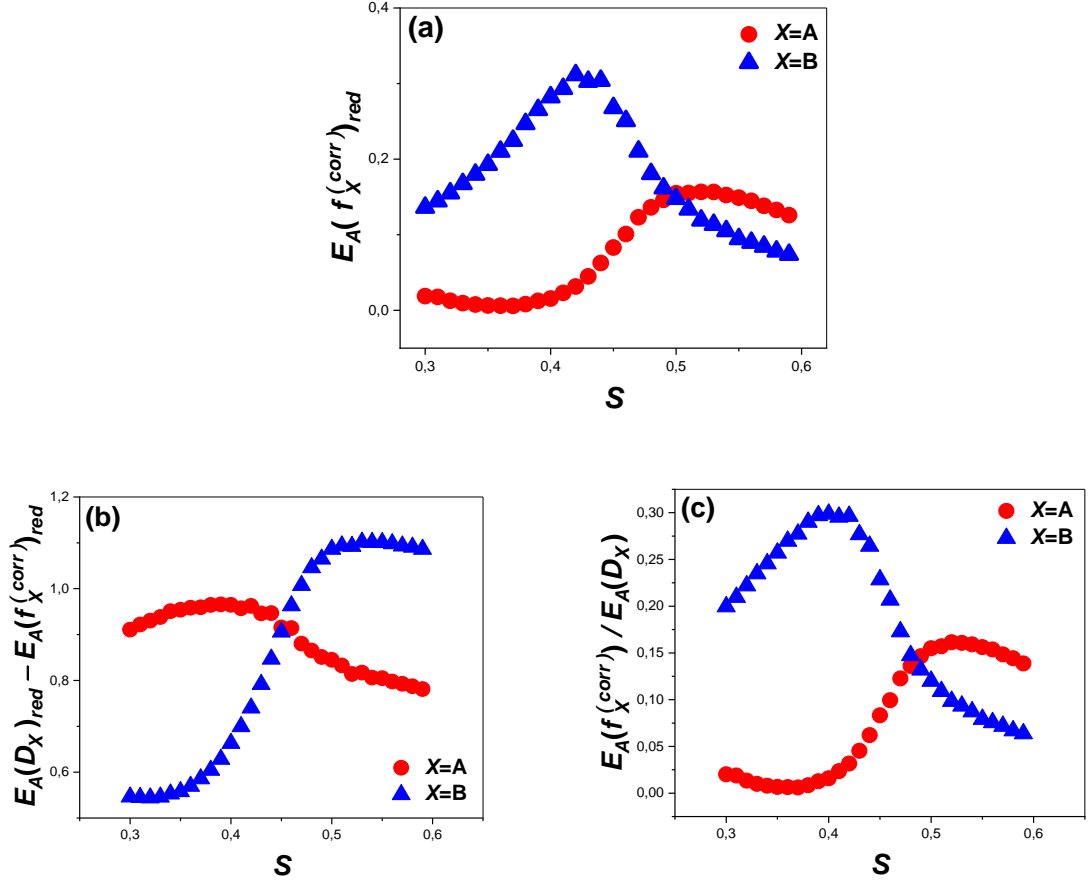

Fig.15. $S$-dependence of: (a) $\left(E_A\left(f_A^{(corr)}\right)\right)_{red}$ and $\left(E_A\left(f_B^{(corr)}\right)\right)_{red}$; (b) the differences $\left(E_A(D_A)\right)_{red} - \left(E_A\left(f_A^{(corr)}\right)\right)_{red}$ and $\left(E_A(D_B)\right)_{red} - \left(E_A\left(f_B^{(corr)}\right)\right)_{red}$; (c) the ratios $E_A\left(f_A^{(corr)}\right)/E_A(D_A)$ and $E_A\left(f_B^{(corr)}\right)/E_A(D_B)$.

According to Eq. (14) the activation energy $E_A\left(f_X^{(corr)}\right)$ additively contributes to the activation energy $E_A(D_X)$ for X-atom tracer-diffusion. The difference $E_A(D_X) - E_A\left(f_X^{(corr)}\right)$ yields, therefore, the part of $E_A(D_X)$ stemming directly from the kinetics of atomic jumps to vacancies. The graphs in Fig. 15c show, in turn, that the contribution of the activation energy $E_A\left(f_X^{(corr)}\right)$ to the total activation energy $E_A(D_X)$ for X-tracer diffusion never exceeded 30%.

### 4.5.2. Analysis of $D_X(T, S)$ in terms of atomic jump frequencies, atom-vacancy pair-correlations and average migration barriers.

The pure effect of $w_{\mu \to \nu}^{(X)}$ on the diffusivities is manifested by the values of $D_X(S,T)$ evaluated with Eq.(13) divided by the corresponding values of $f_X^{(corr)}(S,T)$ evaluated independently with Eq.(15). Elucidation of the atomistic origin of the observed features of A- and B-atom diffusivities follows, in turn, from the analysis of the interrelations between atom-vacancy pair-correlations $(C_{XV}^{(\mu\nu)})$, average migration energies $(\langle E_{X,i(\mu) \to j(\nu)}^{(m)} \rangle)$ and the atomic-jump frequencies $(w_{\mu \to \nu}^{(X)})$.

Fig.16 presents two sequences of the isotherms $\left[ D_X / f_X^{(corr)} \right](S,T)$, $w_{\mu \to \nu}^{(X)}(S,T)$, $C_{XV}^{(\mu\nu)}(S,T)$ and $\langle E_{X,\mu \to \nu}^{(m)} \rangle(S,T)$ corresponding to $T_{red} = 0{,}47$ and $T_{red} = 0.78$.

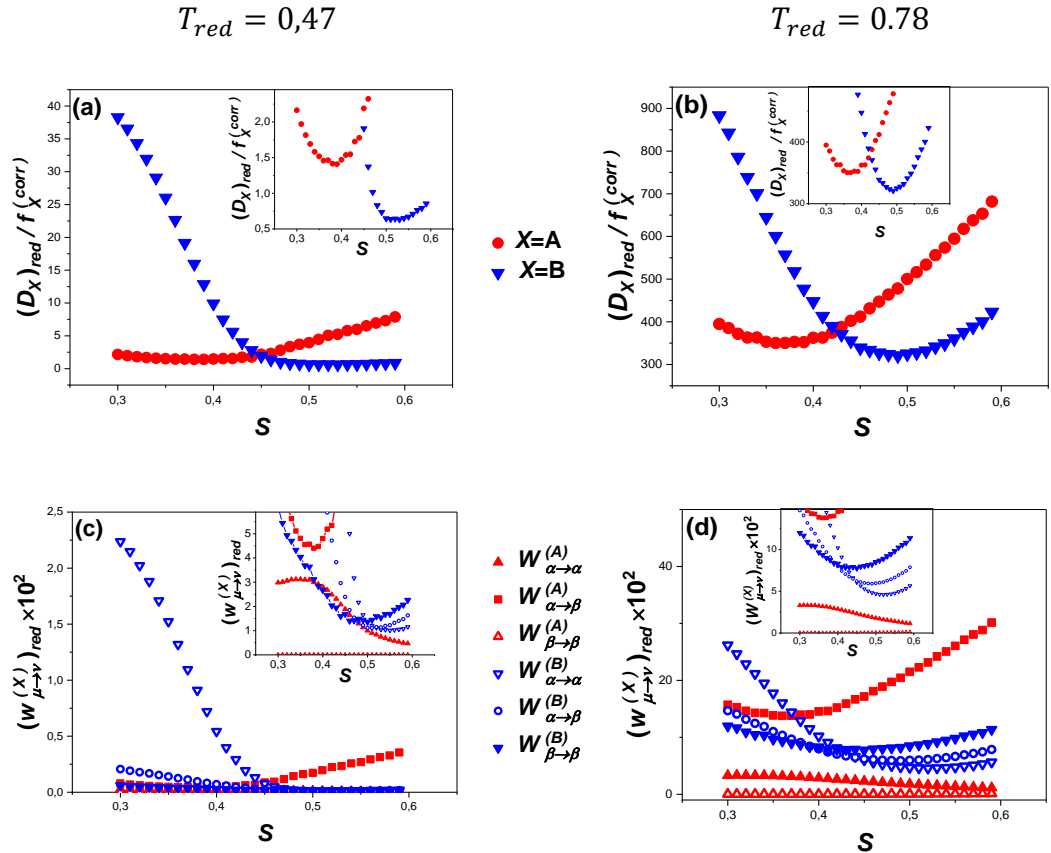

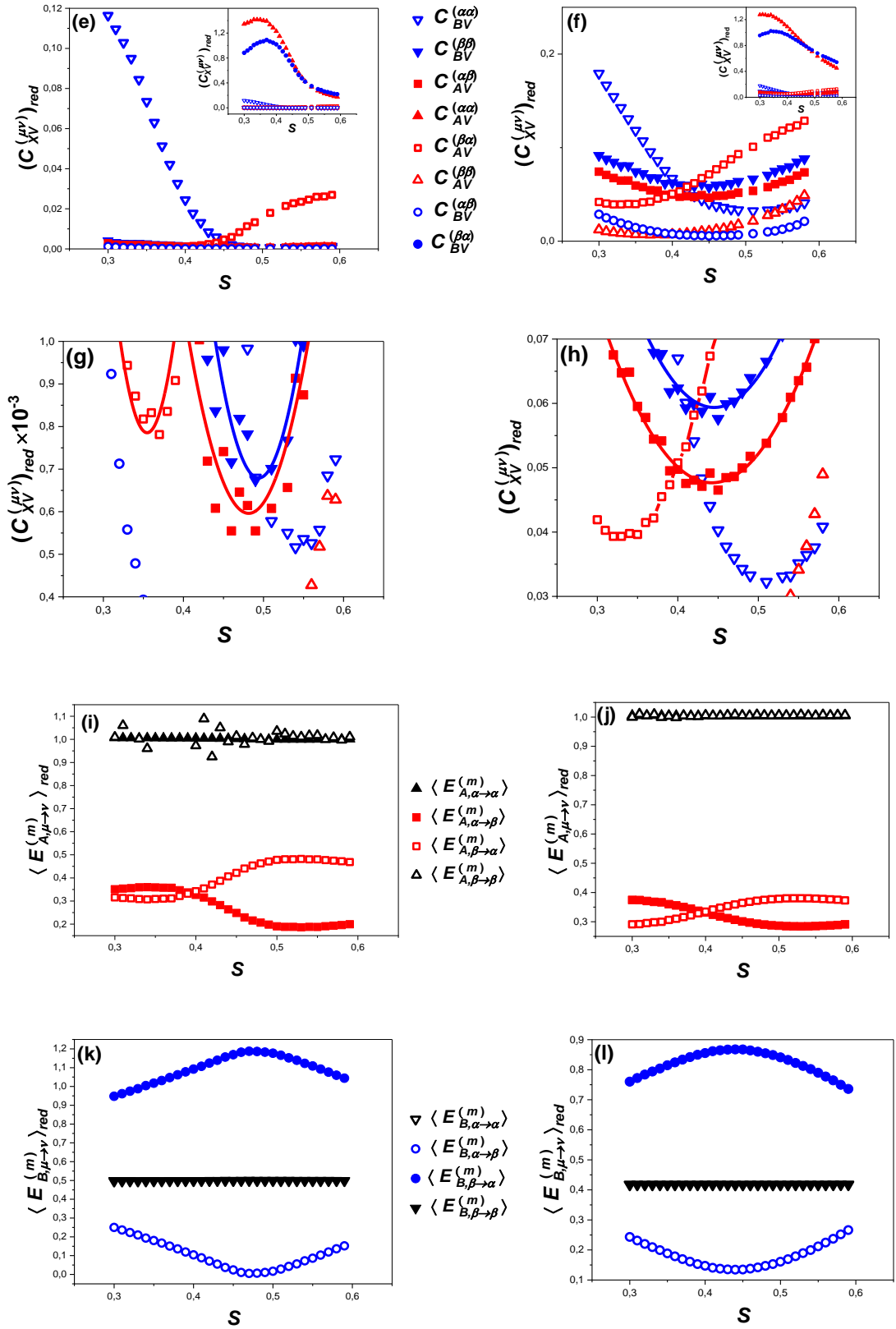

Fig. 16. Isothermal $S$-dependence at $T_{red} = 0.47$ and $T_{red} = 0.78$ of the reduced values of:

$D_X/f_X^{(corr)}$ (a),(b); $w_{\mu \to \nu}^{(X)}$ (c),(d); $C_{XV}^{(\mu \nu)}$ (e),(f),(g),(h); $\langle E_{X,\mu \to \nu}^{(m)} \rangle$ (i),(j),(k),(l). The graphs (g) and (h) show $C_{XV}^{(\mu \nu)}(S)$ in a scale visualizing the minima discussed. The solid lines are for eye guidance only.

The graphs in Figs. 16(a) – (d) clearly indicate that:

(i)     The shapes of $D_X(S)$ (Fig. 6) and $\left[ D_X/f_X^{(corr)} \right](S)$ are qualitatively similar. The division by $f_X^{(corr)}$ slightly shifted the minima of $D_A(S)$ and $D_B(S)$ towards lower and higher values of S, respectively.

(ii)    Only nn intersublattice A-atom jumps are effectively active and the $S$-dependence of their frequencies $w_{\alpha \to \beta}^{(A)}$ definitely controls the shape of $\left[ (D_A)_{red}/f_A^{(corr)} \right](S)$. In particular, the minimum value of $(D_A)_{red}/f_A^{(corr)}$ and its substantial increase in the range of $S > 0.4$ result, from a similar behaviour of $w_{\alpha \to \beta}^{(A)}$.

(iii)   B-atom diffusion proceeds via all three kinds of nn and nnn B-atom jumps, however, while the nnn $\alpha \leftrightarrow \alpha$ jumps and the much less frequent nn $\alpha \leftrightarrow \beta$ ones definitely dominate in the range of $S < 0.5$, the nnn $\beta \leftrightarrow \beta$ jumps are the most frequent in the range of $S \geq 0.5$. Therefore, while the strong increase of $D_B$ at $S < 0.5$ is due to the strong increase of $w_{\alpha \to \alpha}^{(B)}$, the minimum of $D_B(S)$ at $S \approx 0.5$ actually results from the minimum of $w_{\beta \to \beta}^{(B)}$.

(iv)    Increasing the temperature causes a general increase of all the atomic-jump frequencies and shifts the minima of $w_{\mu \to \nu}^{(X)}(S)$ towards lower values of S. Mutual relationships between the values of $w_{\mu \to \nu}^{(X)}$ remain unaffected.

The graphs in Figs. 16 (c) - (h) illustrate the crucial effect of the equilibrium atomic configuration of the system parameterized by the atom-vacancy pair-correlations $\left\{ C_{XV}^{(\mu \nu)} \right\}$ and of the average migration energies $\left\{ \langle E_{X,\mu \to \nu}^{(m)} \rangle \right\}$ on the corresponding atomic-jump frequencies $\left\{ w_{\mu \to \nu}^{(X)} \right\}$ and finally, on the $S$-dependence of the tracer diffusivities $D_A$ and $D_B$.

It is concluded that:

(i)     The fast increase of the B-atom diffusivity in B-rich binaries is mainly due to the fast increase of the B-antisite – $\alpha$-vacancy correlations $C_{BV}^{(\alpha \alpha)}$ which, together with a low and almost constant value of the corresponding average migration energy $\langle E_{B,\alpha \to \alpha}^{(m)} \rangle$, yields a very strong increase of the corresponding atomic jump frequency $w_{\alpha \to \alpha}^{(B)}$. The process is additionally supported by a much weaker increase of both $C_{BV}^{(\beta \alpha)}$ and $C_{BV}^{(\alpha \beta)}$ which, when accompanied with a decrease of initially high $\langle E_{B,\beta \to \alpha}^{(m)} \rangle$, yields an increase of $w_{\alpha \to \beta}^{(B)}$.

(ii)    An increase of the A-atom diffusivity in the A-rich binaries is due to a simultaneous increase of both $C_{AV}^{(\beta \alpha)}$ and $C_{AV}^{(\alpha \beta)}$ occurring at almost constant values of $\langle E_{B,\beta \to \alpha}^{(m)} \rangle$ and $\langle E_{B,\alpha \to \beta}^{(m)} \rangle$.

(iii)   The minimum of $D_B(S)$ and of the corresponding jump frequency $w_{\beta \to \beta}^{(B)}$ observed

at $S \approx 0.5$ clearly coincides with the minimum of the B-atom-vacancy pair correlation $C_{BV}^{(\beta\beta)}$

(iv)  The atomistic origin of the minimum of the A-atom diffusivity $D_A(S)$ stemming from the minimum of $w_{\alpha\to\beta}^{(A)}(S)$ observed at $S \approx 0.4$ is more complex. The single minimum of $w_{\alpha\to\beta}^{(A)}(S)$ results from an interplay between the definitely non-coinciding minima of $C_{AV}^{(\alpha\beta)}(S)$ and $C_{AV}^{(\beta\alpha)}(S)$ (Fig. 15g,h) and the $S$-dependence of the average migration energies $\langle E_{A,\alpha\to\beta}^{(m)} \rangle$ and $\langle E_{A,\beta\to\alpha}^{(m)} \rangle$. The position of the minimum $w_{\alpha\to\beta}^{(A)}(S)$ almost coincides with a point where the isotherms $\langle E_{A,\alpha\to\beta}^{(m)} \rangle(S)$ and $\langle E_{A,\beta\to\alpha}^{(m)} \rangle(S)$ intersect and which, according to Eq.(12), marks the composition $S$ where the A-atom jumps to a nn vacancy generates no change of the system configurational energy.

### 4.5.3.    Composition dependence of the tracer correlation factors

Low values of $f_X^{(corr)}$ mean that the effective distance travelled by tracer atoms is reduced despite the execution of elementary jumps to vacancies – i.e. a great number of jumps are reversed. It is expected that the probability for the reversals of an X-atom $i(\mu) \to j(\nu)$ jumps – i.e. of X-atom oscillations between the sites $i(\mu)$ and $j(\nu)$, is the higher, the higher is the difference between the corresponding migration energies $\Delta E_{X:i(\mu)\leftrightarrow j(\nu)}^{(m)} = \left| E_{X:i(\mu)\to j(\nu)}^{(m)} - E_{X:i(\nu)\to j(\mu)}^{(m)} \right|$. Fig. 17 clearly illustrates the occurrence of the effect in the simulated A-B binaries.

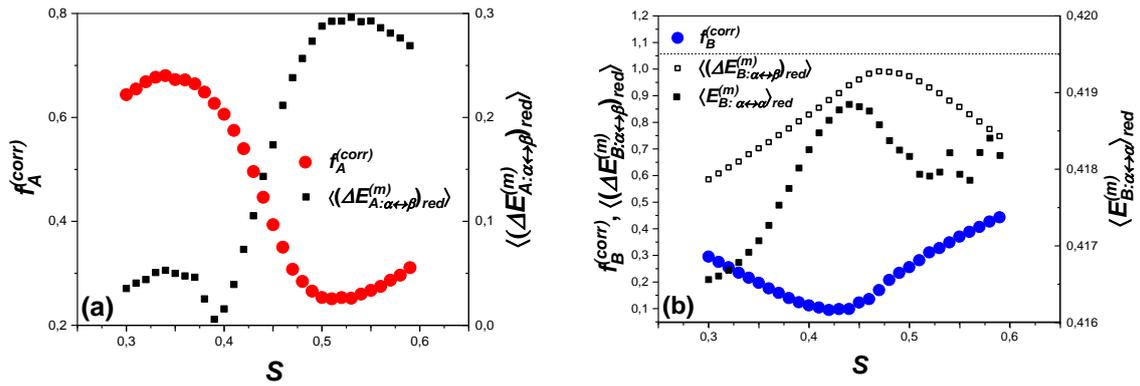

Fig.17. Isothermal S-dependence of the tracer correlation factors $f_X^{(corr)}$ and of the averaged values of $\Delta E_{X:i(\mu)\leftrightarrow j(\nu)}^{(m)}$ corresponding to the dominating X-atom jumps at $T_{red} = 0.47$: (a) $f_A^{(corr)}$ and $\langle \Delta E_{A:i(\alpha)\leftrightarrow j(\beta)}^{(m)} \rangle$; (b) $f_B^{(corr)}$, $\langle \Delta E_{B:i(\alpha)\leftrightarrow j(\alpha)}^{(m)} \rangle$ and $\langle \Delta E_{B:i(\alpha)\leftrightarrow j(\beta)}^{(m)} \rangle$. The dashed line in (b) marks the value of $\left( E_{nnn,bar}^{+}(B) \right)_{red}$.

In the case of A-atoms which migrate predominantly via nn intersublattice jumps the $S$-position of the minimum of the correlation factor $f_A^{(corr)}$ coincides with the maximum of $\langle \Delta E_{A:\alpha\leftrightarrow\beta}^{(m)} \rangle$.

The correlation factor $f_B^{(corr)}$ shows a minimum at $S \approx 0.43$ – i.e. in the area where both nn $\alpha \leftrightarrow \beta$ and nnn $\alpha \rightarrow \alpha$ B-atom jumps are almost equally active and, therefore, both contribute to the effect. The shift between the positions of the minimum $f_B^{(corr)}$ and maximum $\langle \Delta E_{A:\alpha \leftrightarrow \beta}^{(m)} \rangle$ is probably due to the nnn $\alpha \rightarrow \alpha$ jumps. Their average migration barriers are lower than $E_{nnn,bar}^+(B)$ which, in view of Eq.(12), means that the atoms jump most often to positions corresponding to lower configurational energy. An increase and maximum observed on $\langle E_{B:\alpha \leftrightarrow \alpha}^{(m)} \rangle (S)$ (Fig. 17b) suggests that contribution of jumps with a higher migration energy increases and this means more cases of oscillations.

## 5. Discussion

### 5.1. General remarks

The effect of composition and temperature on a steady-state vacancy-mediated tracer diffusion of components was simulated in a B2-ordering A-B binary showing the tendency for TDD. The system was modelled with Ising-type nn ($V_{XY}^{(1)}$) and nnn ($V_{XY}^{(2)}$) pair interactions between atoms *and vacancies* (X,Y = A,B,V) and with migration-barrier parameters $E_{nn,bar}^+(X)$, $E_{nnn,bar}^+(X)$ controlling the heights of the migration barriers encountered by the jumping atoms. The migration-barrier parameters assigned to A- and B-atom jumps to nn and nnn vacancies were evaluated with reference to some *ab initio* calculations concerning Ni-Al [22] and thus, to that extent, the simulated system might be considered as resembling that real one. According to the applied parameterization (Fig. 4, Eq.(12)) the resulting migration barriers were partially dependent on local configurations of the jumping atoms.

Tracer diffusion running in bcc supercells with equilibrium vacancy concentration, as well as with equilibrium atomic and vacancy configuration was simulated by means of a rigid-lattice KMC algorithm. The equilibrium states of the supercells were generated by means of an SGCMC algorithm applied to the Schapink model [24] of phase equilibria in the ternary A-B-V bcc lattice gas.

The main interest was focused on the effect of the tendency for TDD; in particular, on the origin of the 'V'-shapes of the diffusivity isotherms and of the strong enhancement of the B-atom diffusion in the B-rich binaries.

Because the applied rigid-lattice approximation does not allow for the loss of the bcc structure, its stability (e.g. melting), or definite rearrangements of atomic and vacancy configuration (e.g. the formation of phases mimicking $Al_3Ni_2$ or $Al_3Ni_5$ which neighbour β-NiAl in the Ni-Al system [39]) were beyond the performed MC simulations.

### 5.2. Reliability of the model

#### 5.2.1. Ising-type Hamiltonian and migration barriers

As clearly stated earlier, the evaluation of both the values of the pair-interaction parameters ($V_{XY}^{(1)}$ and $V_{XY}^{(2)}$ (X,Y=A,B,V)) and migration-barrier parameters ($E_{nn,bar}^+(X)$ and $E_{nnn,bar}^+(X)$) of the simulated A-B system was mostly arbitrary. Nevertheless,

- The values of $V_{XY}^{(1)}$ and $V_{XY}^{(2)}$ (Table 1) implemented in the Schapink model of the equilibrium atomic configuration [24] reproduced the basic 'TDD' properties of the A-B system (Fig. 9). This result justifies not only the (arbitrary) choice of the values of pair interactions, but also the validity of the Schapink model and its quite specific concept of equilibrium configuration. In view of the technical simplicity of the implementation of the method with SGCMC simulations finally yielding atomic configurations equilibrated simultaneously with respect to vacancy concentration and the degree of chemical order, the result is of importance for the development of this sort of modelling.

- Evaluation of $E_{nn,bar}^{+}(X)$ and $E_{nnn,bar}^{+}(X)$ (X=A,B) was on one hand done with reference to the *ab initio* calculations, but on the other hand, involved the approximation of assigning values of these parameters exclusively to distances of atomic jumps to vacancies and to the kinds of jumping atoms. In view of Eq. (12) the applied migration-barrier parameters affected much more the effective migration barriers for nnn jumps (maintaining the degree of chemical order) than those for nn intersublattice jumps. The applied values of $E_{nn,bar}^{+}(X)$ yielded large difference between the migration eneries for the A-atom and B-atom nn jumps which considerably reduced the B-atom mobility in A-rich binaries. The nn B-atom jumps were, however, never totally blocked – as was done e.g. in ref. [22]. Comparison of the values of the migration energies evaluated *ab initio* for Ni nnn jumps in Ni-Al (Table 2) and the corresponding parameters applied in the present work (Table 3) suggests that the roughest approximation concerned equal barriers for nnn A-atom jumps within α- and β-sublattices. The net effect of this discrepancy on the relationship between the evaluated diffusivities should, however, be small because the very low β-vacancy concentration definitely hindered the A-atom migration within the β-sublattice (regardless of the height of the related migration barriers).

The low value of $E_{nnn,bar}^{+}(B)$ enhanced the migration of B-antisites within the α-sublattice. Their presence resulted, however, from the equilibrium configuration controlled by the pair-interaction parameters. Qualitatively, the value of $E_{nnn,bar}^{+}(B)$ had no effect on the shape of $D_B(S)$, but controlled the position $S(D_A = D_B)$ of the intersection of $D_A(S)$ and $D_B(S)$ (Fig. 18).

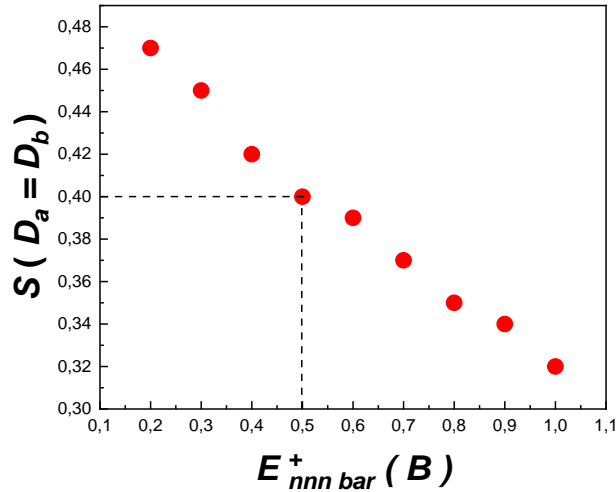

Fig. 18. Effect of $E_{nnn,bar}^{+}(B)$ on the value of $S(D_A = D_B)$ at $T_{red} = 0.47$.

### 5.2.2. Contribution of atomic jumps to nnn vacancies

Occurrence of direct atomic jumps to nnn vacancies, especially in systems showing high B2-ordering energy is considered realistic [40] and their contribution to the vacancy-mediated diffusion in intermetallics was discussed in a number of works concerning the analysis of possible atomistic mechanisms of the phenomenon. Despite an ample number of related studies, most of the works addressing the Ni-Al system (see e.g. [10,14,16,21]) were devoted to Ni self-diffusion in Ni-rich binaries where the nnn jumps of Ni atoms were ruled out due to high migration barriers. Both Ni and Al self-diffusion in B2 Ni-Al was modelled by Xu [22] whose ab-initio calculations – unfortunately, concerning again mainly the stoichiometric and Ni-rich binaries (only one example of an Al-rich alloy was considered) – showed that because of low migration barriers (in contrast to the case of Ni atoms), the nnn jumps of Al atoms cannot be excluded. It was shown that the nnn Al-atom jumps are involved in the proposed 'triple-defect' diffusion mechanism definitely contributing to the process.

In the present work, two features of self-diffusion experimentally observed in Ni-Al were elucidated in terms of the operation of B(Al)-atom jumps to nnn vacancies in the 'triple-defect' A-B (Ni-Al) binaries: (i) the strong increase of $D_{B(Al)}$ in the B(Al)-rich systems resulting from the increase of the frequency of B(Al)-antisite nnn jumps due to both low migration barriers assumed with reference to [22] and the high concentrations of B(Al)-antisites and $\alpha$-vacancies and (ii) the minimum value of $D_B$ at $S = 0{,}5$ resulting from the minimum value of the frequency of B-atom jumps to the nnn $\beta$-vacancies (such jumps were found to dominate at this composition). Regrettably, the results cannot be compared with literature data on the analogous modelling as neither Xu [22], nor, to the authors knowledge, any other authors modelled systematically the concentration dependence of self-diffusion in B(Al)-rich A-B (Ni-Al) binaries.

Finally, it should be mentioned that the results of the present study find support in recent experimental findings obtained for self-diffusion in NiAl by means of X-ray Photon Correlation Spectroscopy [41]. The results suggest a large contribution of atomic jumps in [100] directions, obviously meaning the nnn ones.

### 5.3. Effect of temperature and composition on the system tendency for TDD

According to the definition (see Section 1), the strength of the system tendency for TDD can be measured by the difference between the formation energies for A- and B-antisites. The analysis was performed by applying two alternative parameters measuring the antisite formation energies: (i) $E^{(f)}\left(A^{(\beta)}\right)$ $\left(E^{(f)}\left(B^{(\alpha)}\right)\right)$ equal to differences between average potential energies of A-(B-) atoms on antisite and right positions; (ii) $E^{(f)}\left(A^{(\alpha)} \to V^{\beta}\right)$ $\left(E^{(f)}\left(B^{(\beta)} \to V^{\alpha}\right)\right)$ equal to average increments/decrements of the system configuration energy due to atomic jumps to nn vacancies residing on antisite positions (i.e. due to $A^{(\alpha)} \leftrightarrow V^{\beta}$ and $B^{(\beta)} \leftrightarrow V^{\alpha}$) exchanges). In view of Eq. (12), $E^{(f)}\left(A^{(\alpha)} \to V^{\beta}\right)$ and $E^{(f)}\left(B^{(\beta)} \to V^{\alpha}\right)$ equal the differences $\langle E_{A,\alpha\to\beta}^{(m)}\rangle - \langle E_{A,\beta\to\alpha}^{(m)}\rangle$ and $\langle E_{B,\beta\to\alpha}^{(m)}\rangle - \langle E_{B,\alpha\to\beta}^{(m)}\rangle$ between the average migration energies associated with the nn jumps of A- and B-atoms, respectively.

Fig. 19 shows: (a) isothermal $S$-dependences of $E^{(f)}\left(A^{(\beta)}\right)$, $E^{(f)}\left(B^{(\alpha)}\right)$ and of their differences $\Delta E^{(f)}\left(B^{(\alpha)}, A^{(\beta)}\right) = E^{(f)}\left(B^{(\alpha)}\right) - E^{(f)}\left(A^{(\beta)}\right)$ and (b) isothermal $S$-dependences of $E^{(f)}\left(A^{(\alpha)} \to V^{\beta}\right)$, $E^{(f)}\left(B^{(\beta)} \to V^{\alpha}\right)$ and the difference $\Delta E^{(f)}\left(B^{(\beta)} \to V^{\alpha}, A^{(\alpha)} \to V^{\beta}\right) = E^{(f)}\left(B^{(\beta)} \to V^{\alpha}\right) - E^{(f)}\left(A^{(\alpha)} \to V^{\beta}\right)$.

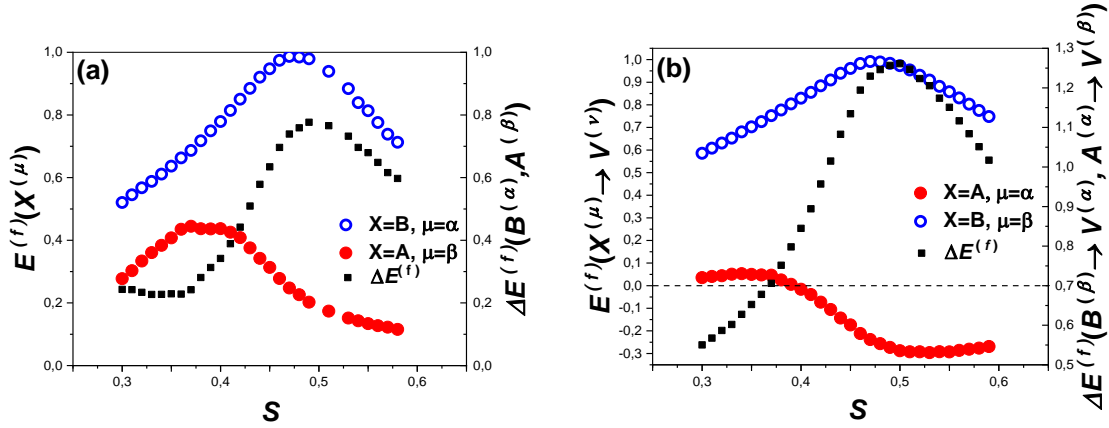

Fig. 19. Isothermal $S$-dependence of (a) $E^{(f)}(A^{(\beta)})$, $E^{(f)}(B^{(\alpha)})$ and $\Delta E^{(f)}(B^{(\alpha)}, A^{(\beta)})$; (b) $E^{(f)}(A^{(\alpha)} \leftrightarrow V^{\beta})$, $E^{(f)}(B^{(\beta)} \leftrightarrow V^{\alpha})$ and $\Delta E^{(f)}(B^{(\beta)} \leftrightarrow V^{\alpha}, A^{(\alpha)} \leftrightarrow V^{\beta})$ at $T_{red} = 0.47$.

Positive values of $\Delta E^{(f)}(B^{(\alpha)}, A^{(\beta)})$ and $\Delta E^{(f)}(B^{(\beta)} \leftrightarrow V^{\alpha}, A^{(\alpha)} \leftrightarrow V^{\beta})$ over all of the range of $S$ indicate the maintenance of the tendency for TDD, whose strength reached, however, a maximum at $S = 0.5$ and continuously decreased when departing from the stoichiometric composition of the system.

It is remarkable that while the values of $E^{(f)}(A^{(\beta)})$ and $E^{(f)}(B^{(\alpha)})$ were positive over all of the explored range of the chemical composition of the system (which guaranteed stability of the B2 superstructure), the energy $E^{(f)}(A^{(\alpha)} \rightarrow V^{\beta})$ was negative for $S > 0.4$ indicating that A-atom-β-vacancy exchanges *decreased* the system configuration energy within this range of the chemical composition.

As was demonstrated in Fig. 9, the ground-state configurations of the A-B binaries showed features typical for a TDD system: the departure from the stoichiometric chemical composition ($S = 0.5$) was compensated *exclusively* by structural A-antisites at $S > 0.5$ and *predominantly* by structural α-vacancies at $S < 0.5$ (few B-antisites started to appear already at $S = 0.5$).
The effect of temperature on the curves corresponding to $T \rightarrow 0$ K (Fig. 9) is illustrated by Fig.20 which displays the same curves together with the analogous equilibrium point-defect concentration isotherms corresponding to $T_{red} = 0.47$.

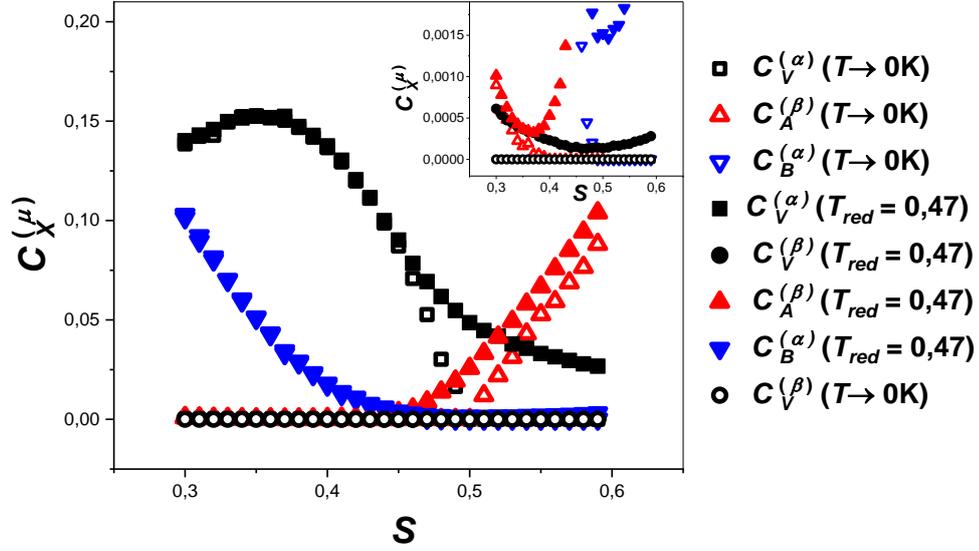

Fig.20. Isothermal S-dependences of the point defect concentrations in the simulated A-B binary at $T \to 0$ K and at $T_{red} = 0{,}47$.

The graphs indicate that the isotherms corresponding to $T > 0$ K conserved the general memory of their shapes at $T \to 0$ K. The temperature effect on equilibrium configurations consisted of the activation of thermal antisites and vacancies causing an increase of the values of $C_V^{(\alpha)}$, $C_V^{(\beta)}$, $C_A^{(\beta)}$ and $C_B^{(\alpha)}$. The following particular temperature-induced features directly influenced tracer diffusivities: (i) monotonic growth of $C_V^{(\alpha)}$ with $S$ decreasing down to $S \approx$ 0.35; (ii) the presence of $\beta$-vacancies (absent at $T \to 0$ K) with concentration $C_V^{(\beta)} \ll C_V^{(\alpha)}$ showing a sharp minimum at $S \approx 0.5$ (see the inset in Fig.19); (iii) replacement of the onsets of the appearance of A-antisites ($S \approx 0.4$) and B-antisites ($S \approx 0.5$) at $T \to 0$ K by minima of $C_A^{(\beta)}(S)$ and $C_B^{(\alpha)}(S)$. Positions of those minima moved with increasing temperature towards lower values of S.

### 5.4. Effect of the system tendency for TDD and its decay on vacancy-mediated atomic migration and tracer diffusion

As the tendency for TDD belongs to the domain of equilibrium thermodynamics, it affected directly the composition dependence of the atom-vacancy correlations $C_{XV}^{(\mu\nu)}$. The principal factors of this influence were: (i) very low values of $C_V^{(\beta)}$ over all the range of $S$ and (ii) continuous increase of $C_V^{(\alpha)}$ with decreasing $S$ stemming from the presence of structural vacancies in the B-rich systems. As a result, the B-rich binaries contained many more vacancies than did the A-rich ones which resulted in a much stronger increase of the values of $C_{BV}^{(\mu\nu)}$ with *decreasing S* with respect to the analogous increase of the values of $C_{AV}^{(\mu\nu)}$ with *increasing S* (Fig. 16 e,f).

The final effect of $C_{XV}^{(\mu\nu)}$ on the atomic-jump frequencies involved the values of the migration energies $\langle E_{X,\mu \to \nu}^{(m)} \rangle$ depending on the parameters $E_{nn,bar}^+(X)$ and $E_{nnn,bar}^+(X)$ (Table 3).

In such terms, the 'V'-shapes of the $D_A(S)$ and $D_B(S)$ isotherms may be discussed by pointing out two aspects: (i) the origin and positions of the minima and (ii) the origin of different slopes of the growth of the diffusivities away from the minima.

### 5.4.1.   Minima of $D_X(S)$

- The minimum of $D_B(S)$ is observed at $S \approx 0.5$ (Fig. 12) where B-atoms migrate predominantly via nnn jumps within β-sublattice (Fig.15c,d). Due to the effect of $f_B^{(corr)}$ it is slightly displaced with respect to the minimum of the corresponding frequency $w_{\beta \to \beta}^{(B)}$ clearly correlated with the minimum of $C_{BV}^{(\beta\beta)}$ which finally follows from the minimum of $C_V^{(\beta)}$ observed at $S \approx 0.5$.

- The minimum of $D_A(S)$ is observed away from the stoichiometric composition at $S \approx 0.43$ (Fig. 12) and similarly, as in the case of $D_B(S)$, the $S$-dependence of $f_A^{(corr)}$ shifts its position away from the minimum of $w_{\alpha \to \beta}^{(A)}$. The minimum of $w_{\alpha \to \beta}^{(A)}$ results, however, from an interplay between the $S$-dependences of different atom-vacancy pair correlations $C_{AV}^{(\alpha\beta)}(S)$ and $C_{AV}^{(\beta\alpha)}(S)$ and the corresponding average migration energies $\langle E_{A,\alpha \to \beta}^{(m)} \rangle (S)$ and $\langle E_{A,\beta \to \alpha}^{(m)} \rangle (S)$ which also differ one from each other and strongly vary with $S$ in the vicinity of the minimum.

In conclusion, the atomistic origins of the minima of $D_A(S)$ and $D_B(S)$ are substantially different and the presented simulation results yield no reasons for their possible coincidence – which, if observed, could be accidental. Besides, it should be noted that as no known experiments on Ni-tracer diffusion in Ni-Al covered the range of $C_{Ni} < 0.47$, the present results may explain the long-time controversy concerning the 'V'-shape of $D_{Ni}(S)$.

### 5.4.2.   Slopes of $D_X(S)$

Enhancement of the A-tracer diffusion and the resulting increase of $D_A$ at $S > 0.43$ clearly followed from the increase of $w_{\alpha \to \beta}^{(A)} = w_{\beta \to \alpha}^{(A)}$ which was due to the increase of the number of A-antisites. The effect found no support from the vacancy concentration which generally *decreased* despite a very small increase of $C_{AV}^{(\beta)}$. As a result, a moderate increase of $C_V^{(\beta\alpha)}$ and a weak increase $C_{AV}^{(\alpha\beta)}$ were observed which combined with constant $\langle E_{A,\alpha \to \beta}^{(m)} \rangle < \langle E_{A,\beta \to \alpha}^{(m)} \rangle$ and yielded an increase of $w_{\alpha \to \beta}^{(A)}$. Despite a very high value of $C_{AV}^{(\alpha\alpha)}$ almost no nnn A-atom jumps occurred because of the very high migration barrier assumed. On the other hand, a very high value of the migration energy $\langle E_{B,\beta \to \alpha}^{(m)} \rangle$ and a very low concentrations of B-antisites and β-vacancies naturally reduced the B-atom activity at $S > 0.5$. Some increase of $D_B$ (whose value was definitely lower than $D_A$) occurred due to an increase of $w_{\beta \to \beta}^{(B)}$ and $w_{\alpha \to \beta}^{(B)}$ following from an increase of $C_V^{(\beta)}$ and consequently of $C_{BV}^{(\beta\beta)}$ and $C_{BV}^{(\alpha\beta)}$ (Fig. 16c-f).

In contrast to the case of A-atom diffusion at $S > 0.43$, the strong enhancement of B-atom diffusion observed at $S < 0.5$ was due to increasing numbers of B-antisites *and α- and β-* *vacancies* and a decrease of $\langle E_{B,\beta\to\alpha}^{(m)}\rangle$ resulting in an increase of the frequencies of all kinds of the B-atom jumps with definite domination of $w_{\alpha\to\alpha}^{(B)}$. The increase of the $\alpha$-vacancy concentration was very strong and occurred in parallel with the decay of the tendency for TDD followed, in turn, by raising the number of B-antisites and thus, by increasing B-antisite-$\alpha$-vacancy correlation $C_{BV}^{(\alpha\alpha)}$. Due to the low migration energy $\langle E_{B,\alpha\to\alpha}^{(m)}\rangle$, the jump frequency $w_{\alpha\to\alpha}^{(B)}$ was also intensively increasing. The also observed growth of $w_{\alpha\to\beta}^{(B)}$ and $w_{\beta\to\beta}^{(B)}$ was caused by the diminishing difference between $\langle E_{B,\alpha\to\beta}^{(m)}\rangle$ and $\langle E_{B,\beta\to\alpha}^{(m)}\rangle$, as well as by growing $C_V^{(\beta)}$.

The intensive increase of $D_B$ started right at $S \approx 0.5$ – i.e. before $D_A$ reached its minimum and in view of the much weaker $S$-dependence of the A-atom diffusivity the value of $D_B$ became finally higher than $D_A$.

### 5.5. Outlook for further investigations

*Ab initio* modelling of Ni-Al yields the virtual B2→A2 'order-disorder' transition point in these binaries close to 6000 K [42] – far above the experimentally observed melting point (inaccessible of course within the rigid-lattice MC simulations). Hence, the reduced temperatures $T_{red}$ corresponding to most of the reported diffusion experiments performed on Ni-Al (usually at $T \approx 1000$ K) do not exceed the level of 0.2. As MC simulations performed at such low temperature are inefficient and yield large uncertainties of the evaluated parameters, the computer experiments are performed much higher in the reduced-temperature scale. In this way, reliable temperature dependences of the parameters of interest are determined. By extrapolating these dependences to the experimental conditions, not only qualitative, but also quantitative correspondence between the simulated and real properties of Ni-Al and other strongly ordered systems might be attained. Such an option seems especially attractive for KMC simulations implemented with *ab-initio* based Hamiltonians – e.g. parameterized with Effective Cluster Interactions (ECI) evaluated within the Cluster Expansion (CE) formalism (as was done in ref.[22]).

### 6. Conclusions

- Vacancy-mediated tracer diffusion of the components of a B2-ordering binary systems A-B showing a tendency for TDD and loosely resembling the Ni-Al compounds was simulated in a wide range of concentration by means of a KMC algorithm. The process was run in crystals with equilibrium configurations and vacancy concentrations generated by a SGCMC algorithm. Features of the temperature and composition dependence of the component diffusivities were elucidated in terms of the frequencies of elementary atomic jumps to nn and nnn vacancies.

- High B(Al)-atom diffusivity in B(Al)-rich binaries was found to be due to enhanced B(Al)-antisite migration via jumps to nnn $\alpha$-vacancies. The diffusivity strongly increased with increasing B(Al)-atom concentration because of the strong increase of both $\alpha$-vacancy and B(Al)-antisite concentrations caused by a gradual decay of the system tendency for TDD.

- The isothermal concentration dependence of the B(Al) atom tracer diffusivity definitely showed a 'V'-shape with a minimum at the stoichiometric composition ($S = 0.5$). The atomistic origins of the shape, as well as of the position of the minimum were explained.

- Although the 'V'-shape was observed also in the case of the isothermal concentration dependence of the A(Ni) atom tracer diffusion, its atomistic origin was different. The minimum was located at $S < 0.5$ – away from the stoichiometric composition. This finding might suggest the reason for the discrepancies between the related experimental results.

## Conflict of Interest

There is no conflict of interest to declare.

## Acknowledgments

This work was supported by the European Community's Seventh Framework Programme (FP7 PEOPLE-2013-IRSES) [grant number EC-GA 612552]; the Polish Ministry of Science and High Education grant number 3135/7. PR/2014/2]; an Endeavour Fellowship and Australian Research Council Discovery Grants schemes [grant number ERF_RDDH_5049_2016] and Polish National Science Center [grant number 2015/16/T/ST3/00501].
The authors are grateful to ACK Cyfronet in Krakow for providing access to Prometheus supercomputer for the purpose of MC simulation.
Thanks are due to Professor Bogdan Sepiol, Faulty of Physics, Vienna University for highly stimulating discussions concerning the correspondence between the present MC simulation results and the conclusions following from his aXPCS experimental results.

The raw/processed data required to reproduce the presented findings cannot be shared at this time due to technical or time limitations.

## References:

[1]    R.J. Wasilewski, Structure Defects in CsCl Intermetallic Compouds. I. Theory, J. Phys. Chem. Solids, 29 (1968) 39-49. https://doi.org/10.1016/0022-3697(68)90252-7

[2]    H. Bakker, N. A. Stolwijk, M. A. Hoetjeseijkel, Diffusion Kinetics and Isotope Effects for Atomic Migration via Divacancies and Triple Defects in yhe Cscl (B2) Structure, Philos.Mag. A, 43 (1981) 251-264. https://doi.org/10.1080/01418618108239405

[3]    R. Smoluchowski, H. Burgess, Vacancies and Diffusion in NiAl,  Phys. Rev., 76 (1949) 309-310. https://doi.org/10.1103/PhysRev.76.309

[4]    St. Frank, S.V. Divinski, U. Sodervall, C. Herzig, Ni tracer diffusion in the B2-compound NiAl: influence of temperature and composition,  Acta Mater.  49 (2001) 1399-1411. https://doi.org/10.1016/S1359-6454(01)00037-4

[5]    G.F. Hancock, B. R. McDonnell, Diffusion in the intermetallic compound NiAl,  Phys. Stat. Sol. (A), 4 (1971) 143-150. https://doi.org/10.1002/pssa.2210040115

[6]    A. Lutze-Birk, H. Jacobi, Diffusion of [114m]In in NiAl,  Scr. Metall. 9 (1975) 761-765.
       https://doi.org/10.1016/0036-9748(75)90236-7

[7]    A. Paul, A.A. Kodentsov, F. Van Loo, On diffusion in the β-NiAl phase,  J.Alloys
       Comp., 403 (2005) 147-153. https://doi.org/10.1016/j.jallcom.2005.04.194

[8]    Y. Minamino, Y. Koizumi, Y. Inui, In Diffusion in B2-Type Ordered NiAl
       Intermetallic Compound, Defect and Diffusion Forum, 94 (2001) 517-522.
       https://doi.org/10.4028/www.scientific.net/DDF.194-199.517

[9]    T. Hughes, E.P. Lautenschlager, J.B. Cohen, J.O. Brittain.  X-Ray Diffraction
       Investigation of β′-NiAlAlloys,  J. Appl.Phys. 42 (1971) 3705-3716.
       https://doi.org/10.1063/1.1659674

[10]   R. Krachler, H. Ipser, Triple-defect complexes in the B2 intermetallic compound
       NiAl,  Phys. Rev. B, 70 (2004) 054113-1-7.
       https://doi.org/10.1103/PhysRevB.70.054113

[11]   Y. Mishin, D. Farkas, Atomistic simulation of point defects and diffusion in B2 NiAl
       .2. Diffusion mechanisms, Philos. Mag. A, 75, 187-199, (1997).
       https://doi.org/10.1080/01418619708210290

[12]   Y. Mishin, M. J. Mehl, D.A. Papaconstantopoulos, Embedded-atom potential for B2-
       NiAl, Phys. Rev. B, 65, 224114-1-14 (2002).
       https://doi.org/10.1103/PhysRevB.65.224114

[13]   Y. Mishin, A. Lozovoi, A. Alavi, Evaluation of diffusion mechanisms in NiAl by
       embedded-atom and first-principles calculations, Phys. Rev. B, 67 (2003) 014201-1-9.
       https://doi.org/10.1103/PhysRevB.67.014201

[14]   S. De Bas, D. Farkas, Molecular Dynamics Simulations of Diffusion Mechanisms in
       NiAl,  Acta Mater., 51 (2003)1437-1446. https://doi.org/10.1016/S1359-
       6454(02)00537-2

[15]   G. X. Chen, J. M. Zhang, K. Xu, Self-diffusion of Ni in B2 type intermetallic
       compound NiAl,  J.Alloys Comp., 430 (2007)102-106.
       https://doi.org/10.1016/j.jallcom.2006.04.052

[16]   S. Yu, C. Wang, T. Yu, J. Cai, Self-diffusion in the intermetallic compounds NiAl and
       Ni3Al: An embedded atom method study, Physica B: Condens. Matter, 396
       (2007)138-144.  https://doi.org/10.1016/j.physb.2007.03.026

[17]   K. A. Marino, E. A.Carter, First-principles characterization of Ni diffusion kinetics in
       β-NiAl, Phys. Rev. B, 78 (2008)184105-1-11.
       https://doi.org/10.1103/PhysRevB.78.184105

[18]   K. A. Marino, E. A. Carter, Ni and Al diffusion in Ni-rich NiAl and the effect of Pt
       additions, Intermetallics, 18 (2010) 1470-1479.
       https://doi.org/10.1016/j.intermet.2010.03.044

[19]   A.V. Evteev, E.V. Levchenko, I.V. Belova, G.E. Murch, Molecular dynamics simulation of diffusion in a (110) B2-NiAl film,  Intermetallics, 19 (2011) 848-854. https://doi.org/10.1016/j.intermet.2011.01.010

[20]   Q. Xu, A. Van der Ven, First-principles investigation of migration barriers and point defect complexes in B2–NiAl. Intermetallics, 17 (2009) 319-329. https://doi.org/10.1016/j.intermet.2008.11.007

[21]   Q. Xu, A. Van der Ven, Atomic transport in ordered compounds mediated by local disorder: Diffusion in B2-$Ni_xAl_{1-x}$, Phys. Rev. B, 81 (2010) 064303-1-5,. https://doi.org/10.1103/PhysRevB.81.064303

[22]   Q. Xu, First-Principles Investigation of Thermodynamic and Kinetic Properties in Ti-H System And B2-NiAl Compound: Phase Stability, Point Defect Complexes and Diffusion, PhD dissertation, University of Michigan, 2009. https://pdfs.semanticscholar.org/1349/328ca5eb4a7a86b221d6acaea9d3bbf236e2.pdf

[23]   M. Athenes, P. Bellon, G. Martin, Identification of novel diffusion cycles in B2 ordered phases by Monte Carlo simulation, Philos.Mag. A, 76 (1997) 565-585. https://doi.org/10.1080/01418619708214023

[24]   F.W. Schapink, The distribution of vacancies in ordered alloys of the CsCl-type, Scr. Metall., 3 (1969) 113-116. https://doi.org/10.1016/0036-9748(69)90211-7

[25]    Z. Qin, G.E. Murch, Computer simulation of tracer diffusion in a binary ordered alloy with an equilibrium concentration of vacancies, Philos.Mag.A, 68 (1993) 831-841. https://doi.org/10.1080/01418619308219369

[26]   Z. Qin, G.E. Murch, Computer simulation of chemical diffusion in a binary alloy with an equilibrium concentration of vacancies, Philos.Mag.A, 71 (1995) 323-332. https://doi.org/10.1080/01418619508244359

[27]   K. Binder, I.L. Lebowitz, M.K. Phani, M.H. Kalos, Monte carlo study of the phase diagrams of binary alloys with face centered cubic lattice structure, Acta Metall., 29, 1655-1665, (1981). https://doi.org/10.1016/0001-6160(81)90048-1

[28]   A. Biborski, L. Zosiak, R. Kozubski, R. Sot, V. Pierron-Bohnes, Semi-grand canonical Monte Carlo simulation of ternary bcc lattice-gas decomposition: Vacancy formation correlated with B2 atomic ordering in A-B intermetallics,  Intermetallics, 18 (2010) 2343-2352. https://doi.org/10.1016/j.intermet.2010.08.007

[29]   B. Widom, Some Topics in the Theory of Fluids, J.Chem.Phys., 39 (1963) 2808-2812. https://doi.org/10.1063/1.1734110

[30]   A.B. Bortz, M.H. Kalos, J.L. Lebowitz, A new algorithm for Monte Carlo simulation of Ising spin systems, J. Comput. Phys., 17 (1975) 10-18. https://doi.org/10.1016/0021-9991(75)90060-1

[31]   P. Sowa, R. Kozubski, A. Biborski, E.V. Levchenko, A.V. Evteev, I.V. Belova, G.E. Murch and V. Pierron-Bohnes, Self-diffusion and 'order–order' kinetics in B2-

ordering AB binary systems with a tendency for triple-defect formation: Monte Carlo simulation, Philos. Mag., 93 (2013) 1987–1998. https://doi.org/10.1080/14786435.2012.742591

[32]  P. Sowa, A. Biborski, M. Kozłowski, R. Kozubski, I. V. Belova, G. E. Murch, Atomistic origin of the thermodynamic activation energy for self-diffusion and order-order relaxation in intermetallic compounds I: analytical approach, Philos.Mag., 97 (2017) 1361-1374. https://doi.org/10.1080/14786435.2017.1302101

[33]  N.Q. Lam, A. Kumar, H. Wiedersich, Kinetics of Radiation-Induced Segregation in Ternary Alloys in: H.R. Brager and J.S. Perrin (Eds.), Effects of Radiation on Materials: Eleventh Conference, ASTM STP782, 1982, p.985-1007.

[34]  F. Haider, R. Kozubski, T.A. Abinandanan, Simulation Techniques in: W. Pfeiler (Ed.), Alloy Physics. A comprehensive reference., Wiley, Weinheim 2007 (ISBN: 978-3-527-31321-1), p. 653-706.

[35]  A. Einstein, The motion of elements suspended in static liquids as claimed in the molecular kinetic theory of heat, Ann.Phys. 17 (1905) 549-560; M.A. Islam, Einstein–Smoluchowski Diffusion Equation: A Discussion, Phys.Scr. 70, 120-125, (2004). https://doi.org/10.1088/0031-8949/70/2-3/008

[36]  H. Bakker, Tracer Diffusion in Concentrated Alloys in: G. E. Murch and A. S. Nowick (Eds.) Diffusion in Crystalline Solids, , Academic Press, Orlando, 1984, p. 189-256.

[37]  G.E. Murch, Simulation of Diffusion Kinetics with the Monte Carlo Method in: G. E. Murch and A. S. Nowick (Eds.) Diffusion in Crystalline Solids, , Academic Press, Orlando, 1984, p. 379-427.

[38]  A. Biborski, L.Zosiak, R. Kozubski, V. Pierron-Bohnes, Lattice gas decomposition model for vacancy formation correlated with B2 atomic ordering in intermetallics, Intermetallics, 17 (2009) 46-55. https://doi.org/10.1016/j.intermet.2008.09.010

[39]  T.B. Massalski, Binary alloy phase diagrams. Metals Park, OH: ASME; 1987 p. 142.

[40]  H. Mehrer, Diffusion in Solids, Fundamentals, Methods, Materials, Diffusion-Controlled Processes, Springer Series in Solid-State Sciences, Springer-Verlag Berlin Heidelberg 2007, p.347

[41]  B. Sepiol – unpublished results.

[42]  J. Betlej – unpublished results